 \documentclass[twocolumn,twocolappendix]{aastex631}

\shorttitle{Light Curve of V1674 Her}
\shortauthors{Kato et al.}

\begin{document}

\title{A comprehensive light curve model of the very fast nova V1674 Herculis}


\author[0000-0002-8522-8033]{Mariko Kato}
\affil{Department of Astronomy, Keio University,
Hiyoshi, Kouhoku-ku, Yokohama 223-8521, Japan}
\email{mariko.kato@hc.st.keio.ac.jp}

\author[0000-0002-0884-7404]{Izumi Hachisu}
\affil{Department of Earth Science and Astronomy,
College of Arts and Sciences, The University of Tokyo,
3-8-1 Komaba, Meguro-ku, Tokyo 153-8902, Japan}

\author{Hideyuki Saio}
\affil{Astronomical Institute, Graduate School of Science,
 Tohoku University, Sendai 980-8578, Japan}



.


\begin{abstract}
V1674 Her is one of the fastest novae, of which the very early phase
is well observed including optical rise to the peak over 10 magnitudes. 
We present a full theoretical light curve model of V1674 Her.   
Our $1.35~M_\sun$ white dwarf (WD) model with the mass accretion 
rate of $1\times 10^{-11}~M_\sun$~yr$^{-1}$ explains overall 
properties including a very fast rise and decay of the 
optical $V$ light curve. 
The WD photosphere expands up to $21 ~R_\sun$, thus, a $0.26 ~M_\sun$
companion star orbiting the WD every 3.67 hours, is 
engulfed 2.7 hours after the onset of thermonuclear runaway, 
and appears 5.3 days after that.  
The duration of X-ray flash is only 0.96 hours.
The evolution of the expanding envelope and temporal change of 
the photospheric radius are very consistent with observed 
optical and X-ray modulations with the orbital and spin (501 s) periods.
We confirmed that the decay phase of nova light curve is 
well approximated by a sequence of steady-state envelope solutions.
Using time-stretching method of nova light curves, we obtain
the $V$ band distance modulus of $(m-M)_V= 16.3\pm 0.2$,
and determine the distance to be $d=8.9\pm 1$ kpc 
for the interstellar extinction of $E(B-V)= 0.5 \pm 0.05$. 
\end{abstract}






\section{Introduction}
\label{introduction}

The classical nova V1674 Her (Nova Herculis 2021) was discovered 
at 8.4 mag on UT 2021 June 12.537 by Seiji Ueda (cf. CBET 4976). 
It has been observed from radio, optical, UV, X-ray, to gamma-ray   
\citep{dra21, woo21, lin22, pat22, ori22, sok23, bha24, hab24, qui24}.  
Early observational results are summarized 
in \citet{dra21}, \citet{sok23}, and \citet{bha24}.



One of the remarkable features of V1674 Her is rich observational 
data in the very early phase of the outburst. 
Unfortunately the X-ray flash was missed, but dense optical data toward
maximum were obtained over 10 magnitudes rise \citep{sok23, qui24}.
Such a dense time series in the very early phase is the first obtained 
in classical novae, which enable us to study the pre-maximum phase
of nova outbursts.

V1674 Her is an intermediate polar (IP) \citep{pat22}. 
The orbital period of 3.67 hr (0.1529 days) and  
the spin period of 8.36 min (501 s) of the white dwarf (WD) are 
determined in both the optical and X-ray wavelengths 
\citep{mro21br, dra21, ori22, pat22, bha24, qui24}.  

It is not easy to obtain theoretical nova light curves 
for a very early phase toward the optical maximum 
because of numerical difficulties.  
A Henyey-type evolution code, which is widely used in nova calculations,
meets numerical difficulties when the luminosity becomes close to 
the Eddington luminosity (see Equation (\ref{equation_Edd}) later introduced) 
and the WD envelope expands to a giant star size.  
Many numerical works took some approximation to 
skip such an extended phase \citep[see a review of ][]{kat17palermo}. 
\citet{kat22sha} has first succeeded in calculating 
through one full cycle of a nova outburst on a $1.0 ~M_\sun$ WD 
including the very extended phase.

The optical rising phase had been poorly 
observed especially in fast novae.  We were unable to compare 
our theoretical light curves with observation in 
the very early phase of a nova outburst. A rare exception is 
YZ Ret, in which the X-ray flash was first detected 
\citep{kon22wa, kat22shapjl, kat22shc}. However, the optical rising 
phase of YZ Ret was not observed frequently enough.
 
V1674 Her is the first classical nova in which a sufficient
time series of optical light curve is obtained in the rising phase. 
The very fast rise toward the optical maximum suggests a very
massive WD similar to that in YZ Ret
($\sim 1.35~M_\odot$: \citet{kat22shapjl,kat22shc,hac23k}).
Here, we present our light curve models on a $1.35 ~M_\sun$ WD
and compare them with the observations of V1674 Her. 
This is the first quantitative comparison between theory and observation 
in the rising phase of a nova covering an amplitude of over 10 magnitudes.

Our aim of this paper is to present our shell flash models 
and to examine how well our one-dimensional calculation 
reproduces the main properties of V1674 Her. 
This paper is organized as follows. First we present 
our numerical results of the $1.35~M_\sun$ WD model 
in Section \ref{sec_model}.
We compare our model light curves with observational results of V1674 Her
in Section \ref{sec_v1674her}.
Discussion and conclusion follow in Sections \ref{sec_discussion} 
and \ref{sec_conclusion}, respectively. 
The $V$ band distance modulus of V1674 Her is obtained
with the time-stretching method in Appendix \ref{time_stretching_method}.


\startlongtable
\begin{deluxetable*}{llllllllllllccll}
\tabletypesize{\scriptsize}
\tablecaption{Nova Model
\label{table_models}}
\tablehead{\colhead{Model} &\colhead{$M_{\rm WD}$} &
 \colhead{$\dot M_{\rm acc}$} &
\colhead{C mix\tablenotemark{a}} &
\colhead{$\log T_{\rm WD}$} &
\colhead{$t_{\rm rec}$} &
\colhead{$L_{\rm nuc}^{\rm max}$} &
\colhead{$\log T^{\rm max}$} & \colhead{$M_{\rm ig}$}&
\colhead{$t_{\rm flash}$\tablenotemark{b}}&
\colhead{$t_{\rm peak}$}\\
\colhead{ }&
\colhead{ ($M_\sun$) } &
\colhead{($M_\sun$ yr$^{-1}$)   } &
\colhead{ }&
\colhead{(K) }&
\colhead{(yr)}&
\colhead{($ 10^9 L_\sun $) } &
\colhead{(K) }&
\colhead{($10^{-6}M_\sun$) } &
\colhead{ (hr)}&
\colhead{ (hr)}
}
\startdata
A&1.35 & $1\times 10^{-11}$ & 0.1 & 7.39  &156,000  & 31 &8.33& 1.6&0.96& 7.8\\
B\tablenotemark{c}
 &1.35 & $5\times 10^{-10}$ & 0.1 & 7.61  &1,900  & 5.8 &8.28& 1.0& 1.5 & 29
\enddata
\tablenotetext{a}{We increased carbon mass-fraction by 0.1 and
decreased helium abundance by 0.1 at ignition.
The resultant abundance is $X=0.7$, $Y=0.18$, $X_{\rm C}$ = 0.1,
and $Z=0.02$, here $X_{\rm C}$ is the extra carbon mass fraction
beyond that in $Z$.}
\tablenotetext{b}{Duration of the X-ray flash.
This is denoted by $t_{\rm ML}$
in \citet{kat22shapjl} or by $\tau_X$ in \citet{kat22shc}.}
\tablenotetext{c}{Same as model I in \citet{kat22shapjl,kat22shc}.}
\end{deluxetable*}

\begin{figure*}
\epsscale{1.1}
\plottwo{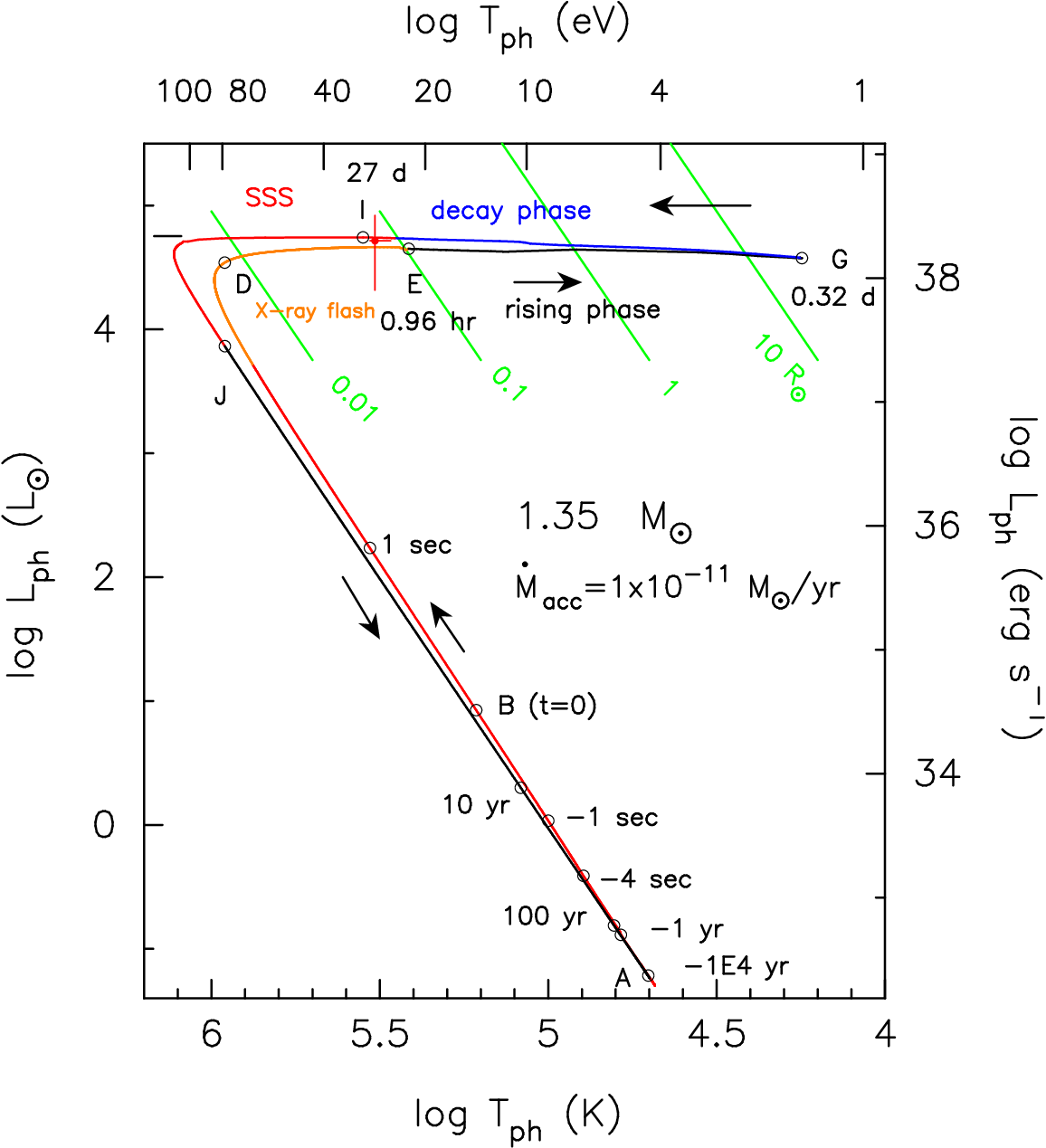}{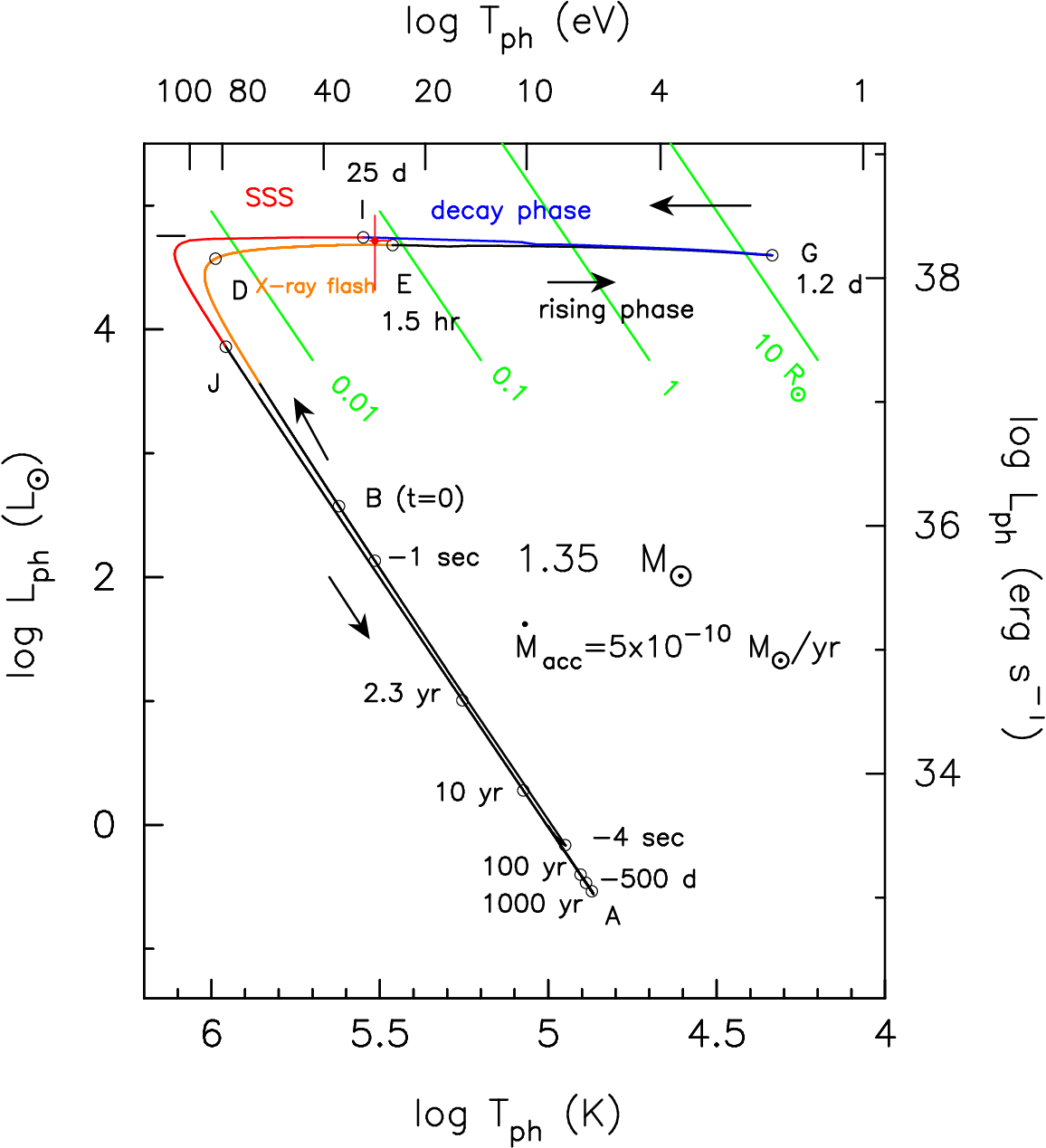}
\caption{
The H-R diagram of one cycle of hydrogen shell flashes for
our outburst models of a $1.35 ~M_\sun$ WD with the mass 
accretion rate of $\dot{M}_{\rm acc}=1\times 10^{-11} ~M_\sun$ yr$^{-1}$
(left panel: Model A)
and  $\dot{M}_{\rm acc}=5\times 10^{-10} ~M_\sun$ yr$^{-1}$ 
(right panel: Model B).
Selected stages during a shell flash are denoted counterclockwise direction 
starting from the bottom of the cycle.  
A: quiescent phase before the shell flash. 
B: the epoch when $L_{\rm nuc}$ reaches maximum ($t=0$).
D: peak of X-ray flash (0.3--1.0 keV). 
E: wind begins to emerge from the photosphere. 
G: the maximum expansion when both the wind mass loss rate and photospheric 
radius reach maximum.
I: The wind mass-loss stops, and the supersoft X-ray phase starts. 
J: The supersoft X-ray luminosity decreases 
to one tenth of the maximum value. 
 The short horizontal line indicates the Eddington luminosity
 (Equation (\ref{equation_Edd})) for the $1.35 ~M_\sun$ WD models.  
The straight green lines show a locus of constant photospheric
radius of $R_{\rm ph}= 10$, 1, 0.1, and $0.01 ~R_\sun$. 
The times at selected stages are indicated before and after stage B.
The filled red circle with error bars indicates the detection 
of YZ Ret in the X-ray flash phase 
\citep{kon22wa}. 
\label{hr}}
\end{figure*}

\section{Nova evolution} \label{sec_model}

\subsection{Numerical Method} \label{sec_method}

We have calculated nova outburst cycles for a 
1.35 $M_\sun$ WD with a mass accretion rate
of $\dot M_{\rm acc}=1\times 10^{-11}~M_\sun$~yr$^{-1}$ 
using our Henyey type evolution code \citep{kat22sha}. 
We stopped mass accretion when the photospheric luminosity
$L_{\rm ph}$ increases to $\log L_{\rm ph}/L_\odot=3.5$ 
and resumes when the luminosity decreases to
less than $\log L_{\rm ph}/L_\odot=2.5$.
We use the OPAL opacity tables \citep{igl96}.
The number of mass grid points is typically 1000--2000 from the center 
of the WD to the photosphere in Henyey calculation part. 
For wind solutions, we adopt 2000--3000 mass grid points. 
The time step changes from the shortest value of 0.01 s 
around the maximum of nuclear energy generation rate 
up to $6 \times 10^9$  s during the quiescent phase. 
When a flash occurs and the photospheric luminosity 
approaches the Eddington limit 
(Equation (\ref{equation_Edd})),
we replaced the hydrostatic radiative surface condition with  
the steady-state optically thick wind solutions \citep{kat94h}.
The epochs of the onset/termination of optically thick winds are
detected with the boundary condition BC1 of \citet{kat94h}.
This boundary condition is designed
for having a smooth connection from the static envelope solution to an
optically thick wind solution \citep[see Table A1 of Appendix A in ][
for details]{kat94h}.
The fitting process of the inner Henyey solution with the steady-state outer
envelope solution sometimes needs many iteration processes as well as
much human time \citep[see ][for details]{kat22sha, kat24M1213}.
Thus, in the wind phase,
the fitting procedure is done at every 10 time steps or more.  

The model parameters and characteristic values 
are listed in Table \ref{table_models}.
From left to right, model name, WD mass, mass accretion rate
in quiescent phase, assumed carbon enhancement in the hydrogen-rich envelope,
temperature at the WD center, 
recurrence period of nova outbursts,  maximum nuclear burning rate,
maximum temperature at the maximum nuclear burning rate,
ignition mass, i.e., mass of the hydrogen-rich envelope at the 
onset of thermonuclear runaway, and
duration of the X-ray flash, i.e., the time when the optically thick winds 
start, and time to the maximum expansion of the photosphere 
(stage G in Figure \ref{hr}) since the onset of thermonuclear runaway.


For comparison, we add a theoretical light curve model of a
$1.35 ~M_\sun$ WD with the mass accretion rate
of $\dot M_{\rm acc}=5\times 10^{-10}~M_\sun$~yr$^{-1}$ (Model B). 
This model is taken from the set of massive WD models 
presented by \citet{kat22shapjl, kat22shc} for the X-ray flash 
in the 2021 outburst of YZ Ret (model I in their list). 
This model shows the shortest optical rising time from the X-ray flash
to the optical peak among the model set of YZ Ret.

We have assumed solar composition for accreting matter
($X=0.7$, $Y=0.28$, and $Z=0.02$). 
Many classical novae, including V1674 Her, 
show heavy element enrichment in ejecta 
\citep[e.g.,][]{geh98tw,hac06kb}. 
%
To mimic such a heavy element enrichment, 
we increased carbon mass fraction of the
hydrogen-rich envelope by 0.1
and decreased helium mass fraction by the same amount at the
beginning of thermonuclear runaway. 

This treatment of element mixing may be too simple to represent actual 
nova outburst; The resultant ejecta abundance is 
different from the estimated values for V1674 Her \citep{hab24}.
However, the aim of this paper is not to present a finely tuned 
model for V1674 Her, but to examine physical properties 
of very fast novae from the theoretical point of view. 
The difference in the chemical composition
will be discussed in Section \ref{sec_composition}

\subsection{One Cycle of Nova outbursts}

Figure \ref{hr} shows one cycle of nova outburst of our models  
in the H-R diagram.  
In the quiescent phase (inter-outburst period), the accreting WD stays
around the bottom of each loop. 
After a thermonuclear runaway starts, the WD moves upward to reach
stage B in the timescale of a second where the energy generation rate 
of nuclear burning reaches maximum, $L_{\rm nuc}=L^{\rm max}_{\rm nuc}$, 
keeping the photospheric radius almost constant. 
We define this epoch as the origin of time, $t=0$. 
The photospheric temperature continuously increases to maximum,
$\log T_{\rm ph}^{\rm max}$ (K)= 5.99.  
With such a high temperature the WD photosphere emits X-ray 
photons, which corresponds to the X-ray flash phase.
After that, the envelope expands and the photospheric temperature
turns to decrease.
Optically thick winds start when the envelope expands and the surface
temperature decreases to $\log T_{\rm ph}$ (K)=5.41 in model A and 
5.46 in model B, both at the open circles labeled E.
The winds possibly self-absorb soft X-rays, thus, we regard this 
point to be the end of X-ray flash.  

When the photospheric radius attains its maximum expansion at epoch G,
the wind mass loss rate also reaches maximum.
The hydrogen-rich envelope quickly loses its mass 
mainly due to wind mass loss.
As the envelope mass is blown in the wind,
the photospheric radius shrinks 
and  the photospheric temperature increases. 
The optically thick winds stop
at epoch I when the photospheric temperature
increases to $\log T_{\rm ph}$ (K) =5.55 in both models A and B.    
The WDs emit X-ray photons and undergo the supersoft X-ray
source (SSS) phase. After that the WD becomes faint and returns to 
the quiescent phase (the bottom of each loop). 

The Eddington luminosity defined by
\begin{equation}
L_{{\rm Edd}} \equiv {4\pi cG{M_{\rm WD}} \over\kappa_{\rm el}}
\label{equation_Edd}
\end{equation}
is indicated with short horizontal bars in the upper-left of
each panels in Figure \ref{hr}, where
$\kappa_{\rm el} = 0.2 (1+X)$ g$^{-1}$ cm$^2$, 
is the electron scattering opacity and $X$ is the hydrogen 
mass fraction. Here we adopt $X=0.55$ from our calculation
in which the hydrogen content has decreased from $X=0.7$ to $X=0.55$.
The photospheric luminosity hardly exceeds the Eddington luminosity
and evolves horizontally before and after the maximum expansion
as already reported in \citet{kov98} and \citet{den13hb}.

Models A and B show very similar evolutions in the H-R diagram, 
except for the positions of stages A, B, and G. 
The mass accretion rate is 
smaller in model A, which results in a fainter quiescent phase. 
Also, model A evolves faster with a larger nuclear burning rate, 
and reaches $L_{\rm nuc}^{\rm max}$ when the photospheric luminosity 
is lower than in model B. A larger envelope mass at ignition in model A makes 
a more extended envelope in the maximum expansion phase.    

\subsection{Evolution of Bolometric Luminosity}

\begin{figure}
\epsscale{1.15}
\plotone{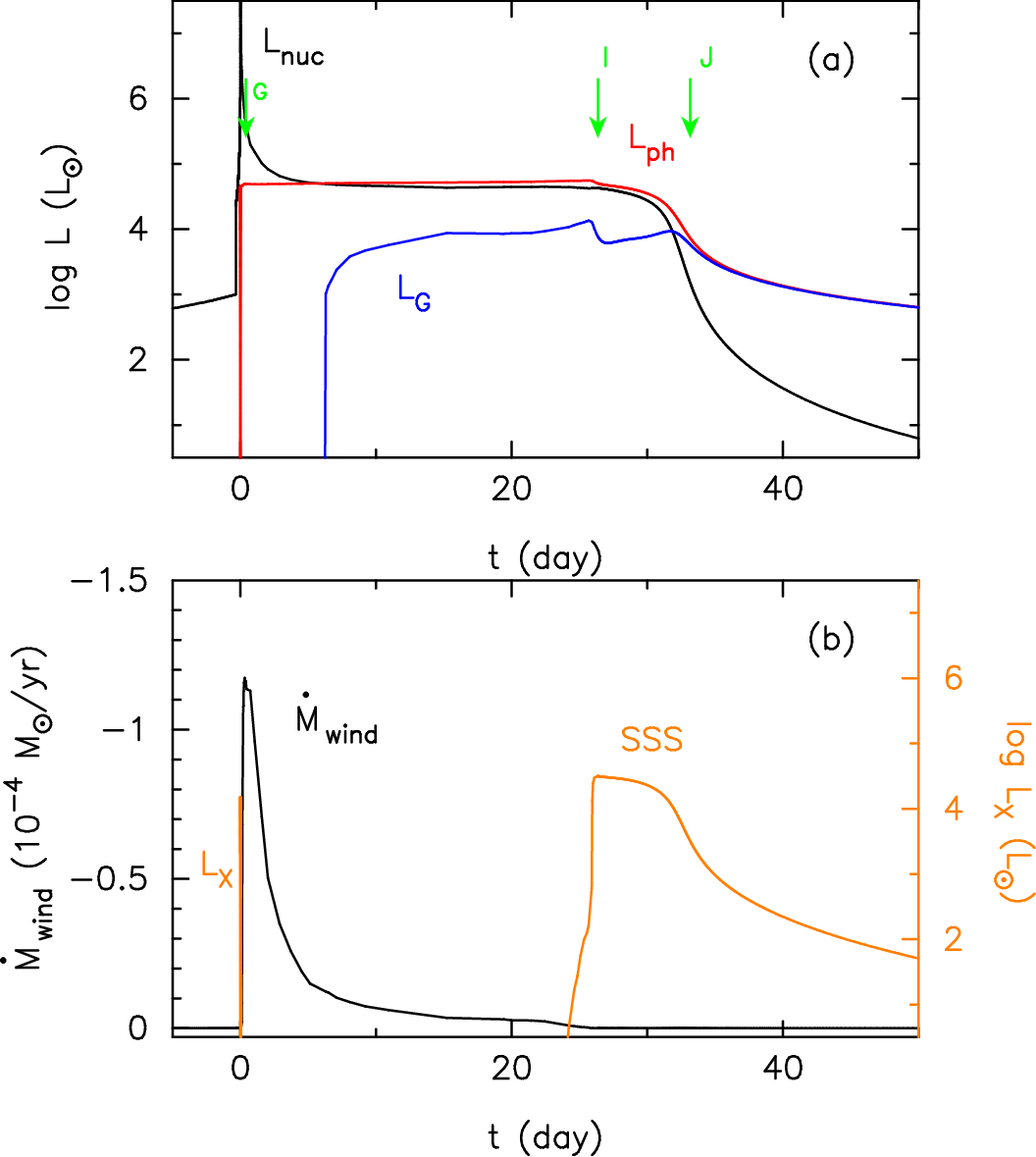}
\caption{(a) The evolutions of the photospheric luminosity, 
$L_{\rm ph}$ (red line), total nuclear burning energy release rate, 
$L_{\rm nuc}$ (black), and total gravitational energy release rate, 
$L_{\rm G}$ (blue) in model A ($1.35 ~M_\sun$ WD with
$\dot{M}_{\rm acc}=1\times 10^{-11} ~M_\sun$ yr$^{-1}$).
We stopped the mass accretion when the luminosity 
increases to $\log L/L_\sun =3.5$ at $t=2.9$ s and restarted
at $ t=72$ day.
Epoch B  in Figure \ref{hr} 
corresponds to $t=0$ and three epochs
G (maximum photospheric expansion), I (end of winds),
and J (supersoft X-ray luminosity decreases to one tenth of its maximum)
are indicated by the downward arrows.
(b) Temporal changes of the wind mass-loss rate (black line) 
and X-ray light curve (orange) in model A. 
}\label{Levo}
\end{figure}

Figure \ref{Levo}a shows the temporal change of 
each energy flux during the thermonuclear runaway in model A. 
Model B shows a very similar evolution and is therefore omitted
(see Figure \ref{TRV.compari} for comparison between model A and model B).
This figure shows the total nuclear energy release rate 
$L_{\rm nuc}=\int \epsilon_{\rm nuc} \delta m$,
the integrated gravitational energy release rate
$L_{\rm G}=\int \epsilon_{\rm g} \delta m$, 
and the emergent bolometric luminosity $L_{\rm ph}$. 
Here, $\int \delta m$ is mass integration in the hydrogen-rich envelope, 
$\epsilon_{\rm nuc}$ and $\epsilon_{\rm g}$ are the energy 
generation rates per unit mass owing to hydrogen nuclear burning 
and gravitational energy release, respectively.
(see Equations (1), (3), and (4) of \citet{hac16sk} for 
their definitions).

In the beginning of the outburst,
the nuclear energy generation rate $L_{\rm nuc}$ is much
larger than the photospheric luminosity $L_{\rm ph}$.
The excess energy ($L_{\rm nuc}-L_{\rm ph})$ is absorbed 
($L_{\rm G} < 0$) and used to expand the
envelope against the gravity.
In other words, a large amount of generated energy is stored 
in the nuclear burning region as gravitational energy (by expansion),
which is released (by shrinkage) later until the end of the shell flash. 

As a result, only a small part of the released energy is 
transferred to the photosphere. 
Because the bottom of the envelope expands, the density
in the burning layer decreases that results in a rapid
decrease in $L_{\rm nuc}$. 
Finally $L_{\rm nuc}$ becomes comparable to $L_{\rm ph}$ on day 4.

The integrated gravitational energy release rate $L_{\rm G}$ increases 
from a negative to a positive value on day 6  
as shown in Figure \ref{Levo}a. 
The absorbed energy is constantly released that amounts to 
10 percent of the photospheric luminosity.  


Figure \ref{Levo}b shows the temporal change of wind mass loss rate 
as well as the X-ray light curve. 
The X-ray flash ($t \sim 0$) lasts very short and hard 
to see; later it will appear in an extended scale (Figure \ref{xrayflash}).
In the SSS phase the X-ray flux rises quickly 
when the optically thick winds become weak and finally 
stop at $\log T_{\rm ph} =5.55$ (on day 27). 
The X-ray luminosity stays at $\log (L_{\rm X}/L_\sun) > 4$ 
during 6 days followed by gradual decline. 


\begin{figure}
\epsscale{1.1}
\plotone{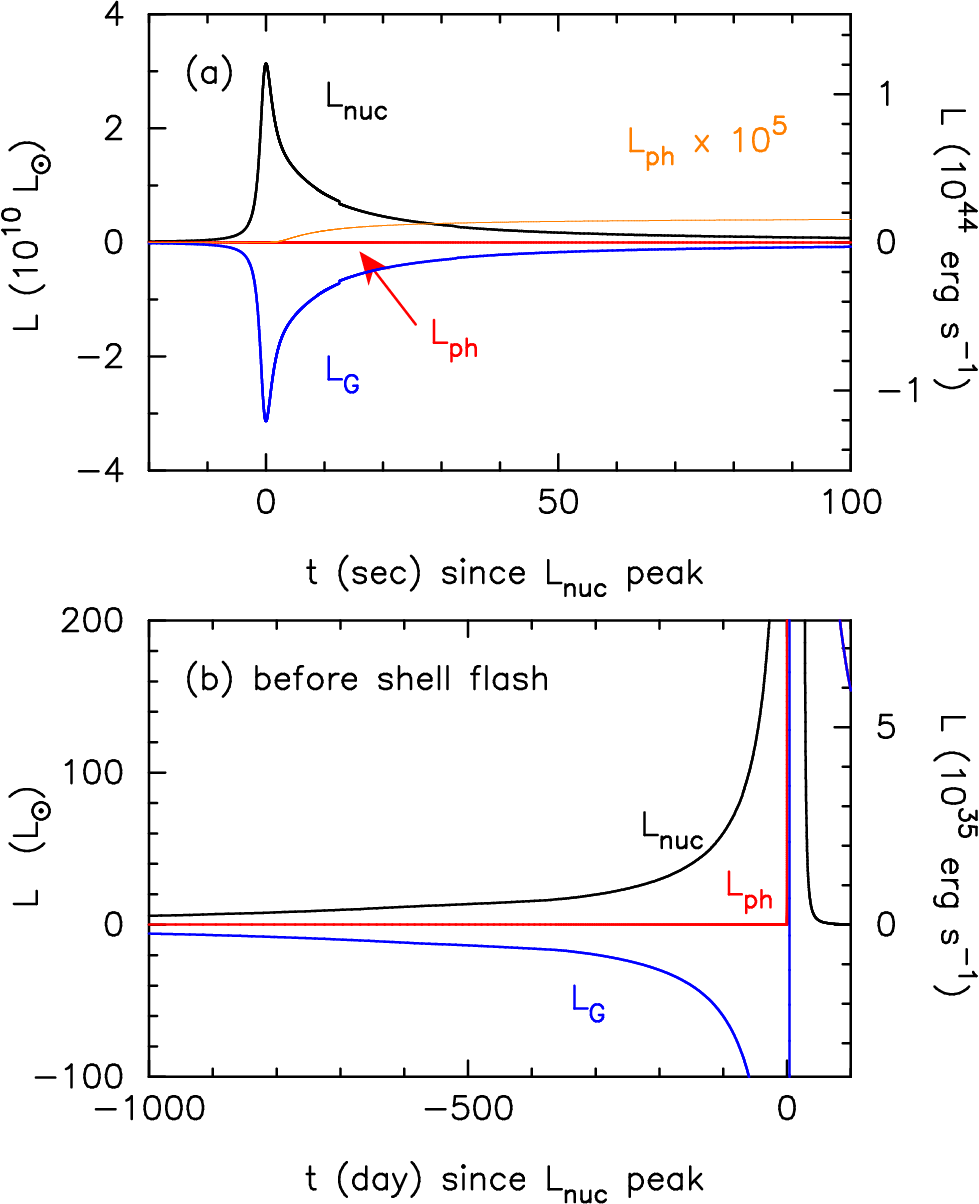}
\caption{
(a) Close-up view of a very early phase of the shell flash
in model A.
The photospheric luminosity, $L_{\rm ph}$,
total nuclear burning energy release rate, $L_{\rm nuc}$,
and total gravitational energy release rate, $L_{\rm G}$ are shown
from $t=-20$ s to $t=100$ s after the onset of the shell flash ($t=0$).
The orange line amplifies the photospheric luminosity (red line) by
$10^5$ times. Note that the luminosity is plotted
against a linear time in units of second.
(b) Same as panel (a), but zoom-out for a pre-outburst phase,
from $t= -1000$ day to $t=100$ day. 
Note that the luminosity is plotted against a linear time
in units of day.
}\label{L}
\end{figure}

Figure \ref{L}(a) shows a close-up view of the energy conservation during
the epoch of thermonuclear runaway in model A.
Note that the luminosity is plotted against a linear time.
The nuclear energy release rate increases in a timescale of
a few seconds to reach the maximum value of 
$L_{\rm nuc}^{\rm max}= 3.13 \times 10^{10}~L_\odot$,
while most of the released nuclear energy is absorbed
(i.e., $L_{\rm nuc}\sim-L_{\rm G}\gg L_{\rm ph}$) to expand the envelope.

Model B shows a very similar temporal change of 
luminosities, whereas $L_{\rm nuc}^{\rm max}$ is 
five times smaller than in model A (see Table \ref{table_models}). 
In both models A and B, the radiative energy flux $L_{\rm ph}$ is 
always smaller than the Eddington luminosity, which is 
already shown in Figure \ref{hr}. 



Figure \ref{L}b shows a zoom of the pre-outburst stage, from 
1000 days before the onset of thermonuclear runaway ($t=0$).  
The nuclear luminosity gradually rises since few years before
the outburst. 
The photospheric luminosity stays at a 
very faint level of $L_{\rm ph} \sim 0.1 ~L_\sun$ 
until just before the onset of thermonuclear runaway 
(See also Figure \ref{hr}).  
Thus, we have no signs of precursor for the outburst significantly
before the outburst.




\subsection{X-ray Flash}

\begin{figure}
\epsscale{0.95}
\plotone{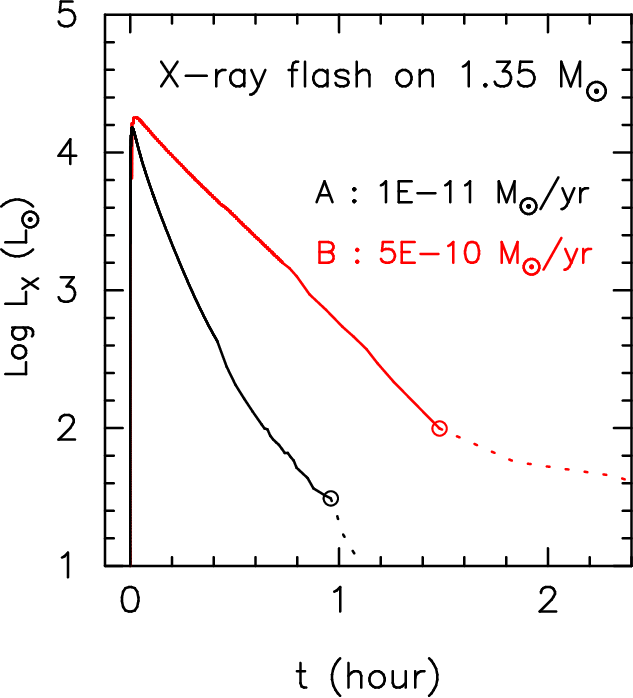}
\caption{The supersoft X-ray light curve (0.3--1.0 keV) 
during the X-ray flash of our 1.35 $M_\odot$  models: 
Model A (black line) and Model B (red line).
The open circles corresponds to stage E when the optically thick 
winds emerge from the photosphere.
The dotted part corresponds to the wind phase, in which X-rays may be 
self-absorbed by wind outside the photosphere.  Here, no absorption
is assumed outside the photosphere. 
}\label{xrayflash}
\end{figure}

Theoretical X-ray light curve (0.3--1.0 keV) is calculated from 
the photospheric temperature and the luminosity of the 
blackbody emission from the WD (Figure \ref{Levo}b). 
Because most of the X-ray photons are emitted at energy below 1.0 keV,
the resultant flux is hardly changed for a different upper bound
higher than 1.0 keV, for example, 0.3--10 keV of the Swift/XRT band.  
We assume no absorption outside the photosphere.

Figure \ref{xrayflash} shows the X-ray light curves during the 
X-ray flashes. 
Model B is reported in \citet{kat22shapjl} but with a 
(0.2--10.0 keV) band (model I in their Figure 3i). 
Model A has a smaller mass accretion rate, 
hence a larger envelope mass at ignition, than model B.  
For a larger ignition mass, $L_{\rm nuc}^{\rm max}$ is larger, which 
results in a rapid evolution of the envelope, and 
the photospheric temperature decreases faster and 
the X-ray luminosity also decreases faster. 
Thus, the duration of X-ray flash is shorter in model A.


\subsection{Internal Structure of the Envelope}

\begin{figure}
\epsscale{1.1}
 \plotone{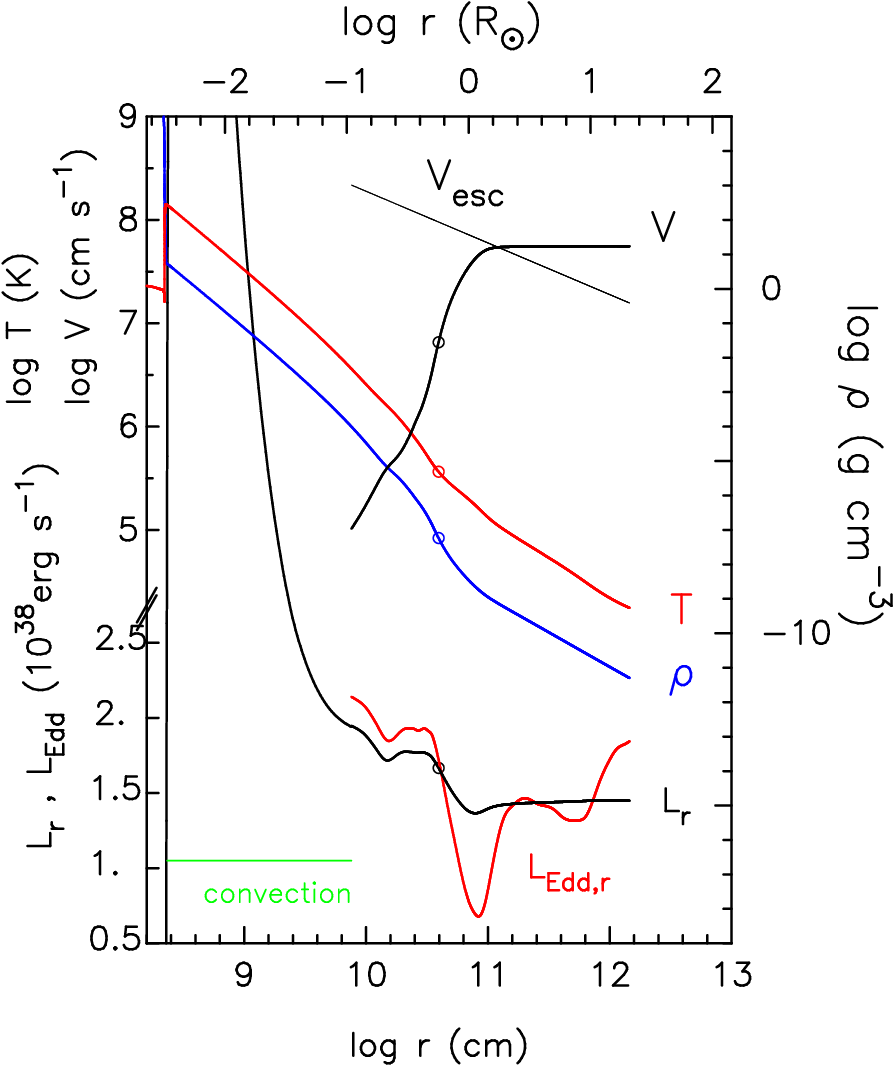}
\caption{The internal structure at stage G (maximum expansion).
The temperature (upper red line labeled $T$), density (blue line, $\rho$), 
photon flux (lower black line, $L_r$), local Eddington luminosity 
(lower red line, $L_{{\rm Edd},r}$), velocity (upper black line, $V$),
and escape velocity (uppermost black line, $V_{\rm esc}$) are shown. 
Convective region is denoted by the horizontal green line.
The open circles denote the critical point \citep{bon52, kat94h} 
of the optically thick wind solution, 
which is located in the acceleration region. 
The right edge of each line corresponds to the photosphere. 
}\label{struc.max}
\end{figure}

Figure \ref{struc.max} shows the internal structure at stage G  
(the maximum expansion of the photosphere), i.e., the
distributions of the temperature, density, velocity, energy flux, and  
local Eddington luminosity in the radiative region, which is defined as
\begin{equation}
L_{{\rm Edd},r} \equiv {4\pi cG{M_{\rm WD}} \over\kappa}.
\label{equation_Edd.local}
\end{equation}
The local Eddington luminosity is inversely proportional to 
the opacity $\kappa$ 
that is a function of the temperature and density.
A small dip in $L_{{\rm Edd},r}$ 
at $\log r$ (cm) $\sim 10.2$ corresponds to
a small peak in the opacity contributed by ionized O and Ne
($\log T$ (K) $\sim 6.2$--6.3), and a large dip
at  $\log r$ (cm) $\sim 10.9$ is caused by
a large Fe peak at $\log T$ (K) $\sim 5.2$
(\citet{igl96}, or Figure 6 in \citet{kat16xflash}).

The velocity quickly increases outward
where the local Eddington luminosity decreases corresponding to
the opacity increase {\it outward}. 
At the same time $L_{r}$ exceeds the local Eddington
luminosity $L_{{\rm Edd},r}$.
The local luminosity $L_r$ decreases outward
in the envelope at/around the critical point. 
This is because the photon
energy is consumed partly to lift the envelope matter up 
against the gravity
(gravitational energy), to heat the envelope (thermal energy),
and to increase the kinetic energy of the winds.
These properties are essentially the same as those reported 
for the $1.0~M_\sun$ WD model \citep{kat22sha}. 

Note that we did not include the convective energy transport 
outside the critical point, 
because the wind velocity is supersonic and therefore
convective eddies cannot turn back.

\subsection{Evolutional Change of Internal Structure}

\begin{figure*}
\epsscale{1.05}
\plotone{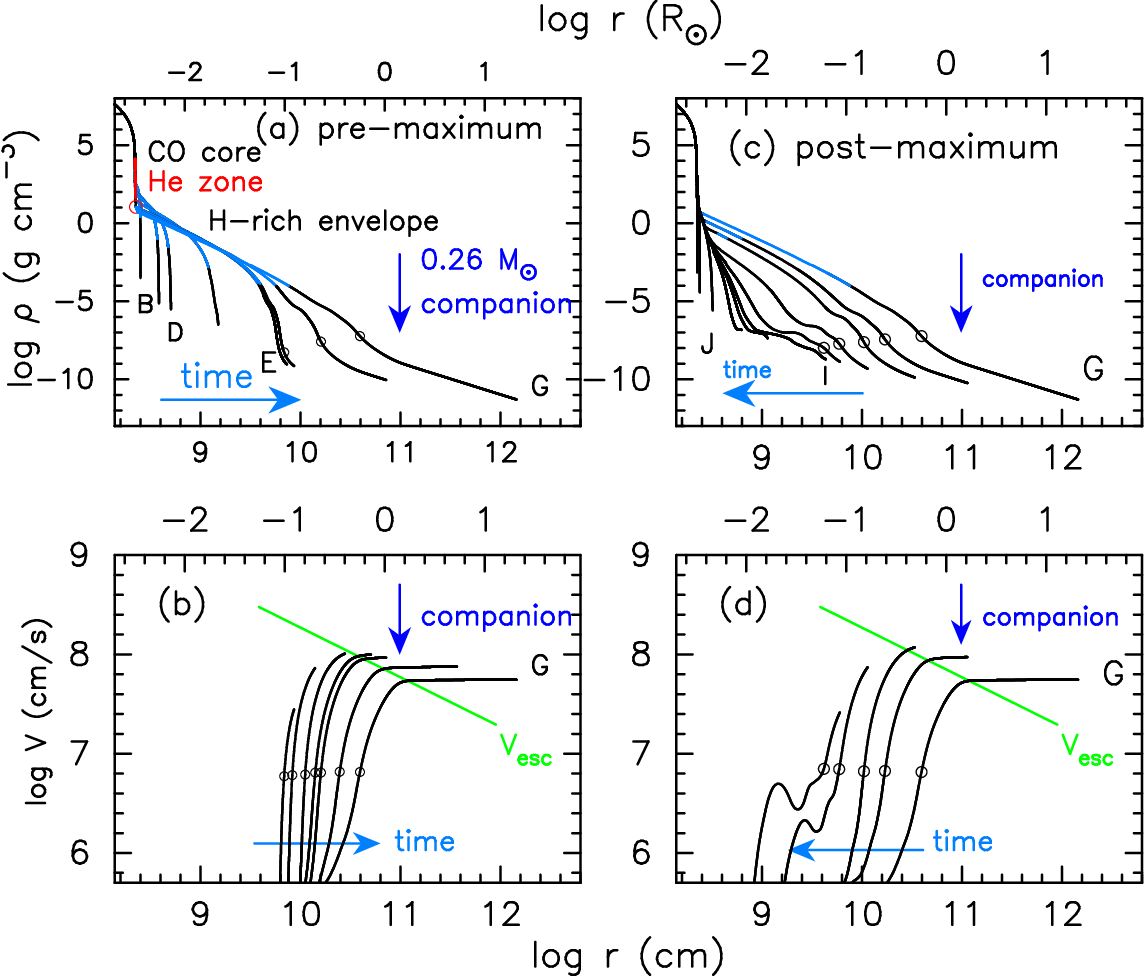}
\caption{
The pre- and post-maximum evolutions of the density and 
velocity profiles of the envelope for some selected stages.
The label B, D, E (winds emerge), G, I, and J correspond to each 
stage in Figure \ref{hr}. The CO core (black) and He-rich layer 
(red) are indicated only for stage  E.
The red open circle denotes the position 
of maximum nuclear burning rate in stage E. 
The convective region of each stage is colored by blue. 
The right edge of each line corresponds to the photosphere. 
The black open circles denote the position of the critical point
of optically thick wind solution \citep{kat94h}. 
The circular orbit of the companion star is shown at $a=1.41~R_\sun$
for $M_{\rm WD}= 1.35 ~M_\sun$, $M_2= 0.26 ~M_\sun$, and 
$P_{\rm orb}= 3.67$ hr.
}\label{rhov.m135}
\end{figure*}

Figure \ref{rhov.m135}a  shows the temporal changes of
the density profiles in the pre-maximum phase.
Before the outburst, the envelope is in plane parallel
structure where the density steeply decreases outward.
Just after the shell flash starts, the convection widely extends
above the nuclear burning zone.
In the early phase, the convective region hardly changes its structure 
and the outer radiative region extends with a steep decreasing density 
profile. 

The optically-thick wind mass-loss starts just after stage E. 
The critical point appears at the photosphere and 
a part of the envelope matter is accelerated up to supersonic. 
In the outer region above the critical point, the density decreases
proportionally to $\rho \propto r^{-2}$ because of the steady state
condition of $4 \pi r^2 \rho v=$ constant with a constant velocity $v$
(see Figure \ref{struc.max}). 

Figure \ref{rhov.m135}b shows the velocity profiles in the rising phase. 
The envelope matter is accelerated in the region around the critical point 
which is inside of the binary orbit. When the matter reaches the 
companion star the matter is already accelerated beyond the 
escape velocity (green line). Further acceleration would be difficult
(see discussion in \citet{kat94h, kat11drag}).  



Figure \ref{rhov.m135}c shows 
the density distribution in the decay phase.
The convective region becomes narrower and disappears. 
The photospheric radius shrinks with time
because the density in the outer envelope decreases.
The optically thick winds stop at stage I. 
The envelope further shrinks as the envelope mass decreases owing to
nuclear burning, and eventually becomes geometrically thin.
At stage J, the structure is in plane parallel.

Figure \ref{rhov.m135}d shows the velocity profiles.
The matter is accelerated in the region where the 
radiative opacity quickly increases outward (See Figure \ref{struc.max}). 
The critical point appears in the center of the accelerated region. 
At the critical point, the velocity reaches the isothermal sound speed 
\citep[e.g. Equation (11) in][]{kat94h}, and the wind becomes supersonic 
outside the critical point. 
The acceleration region moves inward in the $r$ coordinate 
as shown in Figure \ref{rhov.m135}c and d 
by the inward shift of the critical point. 
In the very later phase of the wind phase the velocity profile shows a
small peak inside of the critical point, corresponding to 
the small opacity peak owing to O and Ne at $\log T$ (K) $\sim 6.2$. 
This small bump appears in the later phase of nova wind phase 
of massive WDs. 




\subsection{Optical Light Curve}

We have calculated optical $V$ light curve based on 
the optically thick winds,
assuming that free-free emission dominates the optical/NIR flux
\citep[e.g.,][]{enn77,gal76}. 
The  $V$ luminosity can be simplified as
\begin{equation}
L_{V, \rm ff,wind} = A_{\rm ff} ~{{\dot M^2_{\rm wind}}
\over{v^2_{\rm ph} R_{\rm ph}}}.
\label{free-free_flux_v-band}
\end{equation}
\citep{hac06kb,hac20skhs}.
This $V$ flux represents the luminosity 
of free-free emission from 
optically thin plasma just outside the photosphere.  
We use the wind mass loss rate $\dot{M}_{\rm wind}$, 
the velocity at the photosphere $v_{\rm ph}$, 
and the photospheric radius $R_{\rm ph}$ calculated for
our $1.35 ~M_\sun$ WD model.  

We implicitly assume that novae of the same WD mass and chemical
composition should follow the same $A_{\rm ff}$ light curve.
\citet{hac20skhs} obtained $A_{\rm ff}$ calibrated with the novae
with known chemical composition and distance modulus in the $V$ band,
such as LV Vul.  Because the coefficient $A_{\rm ff}$ is not yet specified
for our adopted chemical composition in V1674 Her, 
we have determined the $A_{\rm ff}$ ($=$ the absolute magnitude
of the light curve) using the 1.35 $M_\sun$ steady state model in 
Figure \ref{individual_v1674_her_v_chemical_logscale_no2} 
that has the closest chemical composition among our database models.
The resultant light curve will be shown in the next section 
(Figure \ref{light.2}).


\section{Comparison with V1674 Her observation}
\label{sec_v1674her}

This section describes how our theoretical model reproduces
the observational properties of the very fast nova V1674 Her.


\subsection{Optical Light Curve}

\begin{figure*}
\epsscale{0.85}
\plotone{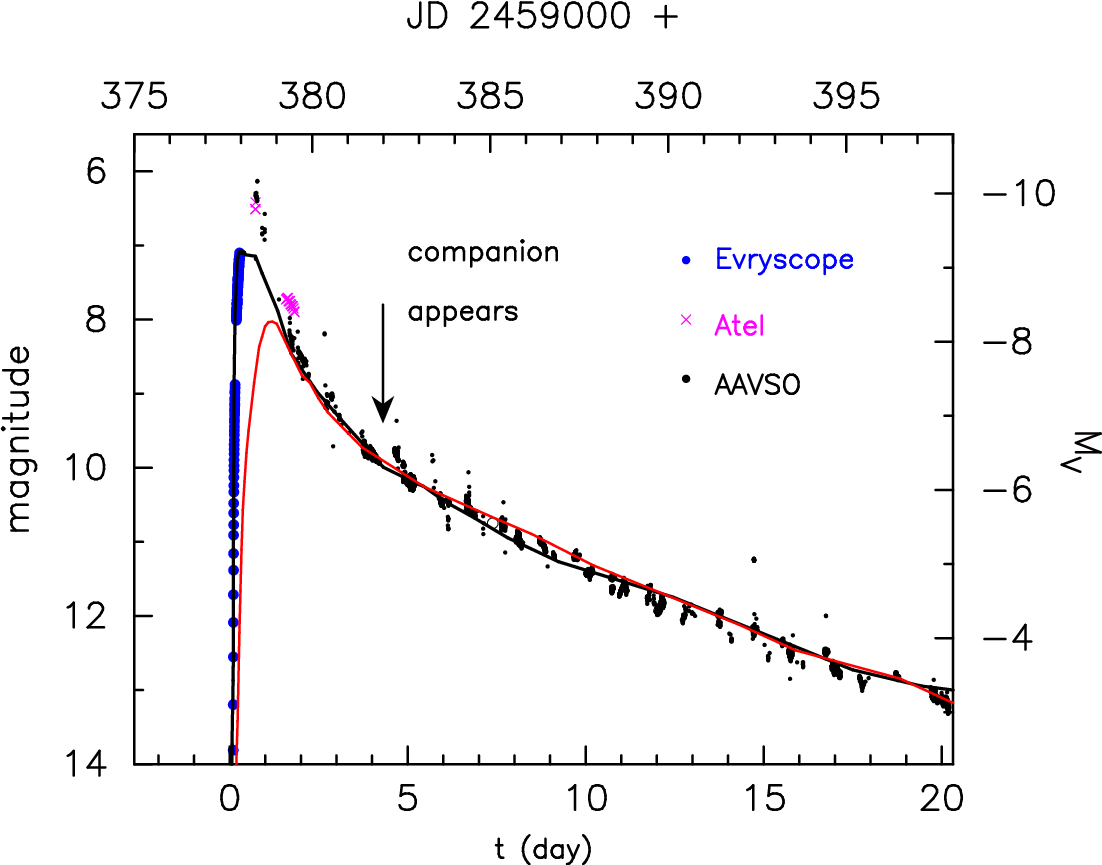}
\caption{
Comparison of our theoretical $V$ light curve with the
observational data in the first 20 days of the V1674 Her 2021 outburst. 
We assumed the origin of time ($t=0$) to be JD 2,459,377.68. 
Blue filled circles: Evryscope g-magnitudes taken from \citet{qui24},   
Black dots: $V$ and visual magnitudes taken from the American Association
of Variable Star Observers (AAVSO). 
Magenta crosses; $CV$ magnitudes taken from ATel \#14710 (E. Aydi et al.).  
The black and red line represent the theoretical free-free emission
$V$ light curves of model A and model B, respectively, with 
the distance modulus in the $V$ band being $\mu_V \equiv (m-M)_V= 16.3$
(the distance and reddening to be 8.9 kpc and $E(B-V)=0.5$, respectively,
see Section \ref{sec_distance}).
The epoch when the outer edge of the companion star re-appears 
from the WD photosphere is  
indicated by the downward arrow for model A. 
}\label{light.2}
\end{figure*}

Figure \ref{light.2} shows observational data 
of $V$, visual, and $g$ magnitudes for the first 20 days. 
The sources of the data are described in the figure caption. 
The JD time is shown in the upper abscissa, and  
the time for theoretical model is in the lower abscissa. 
We fit the origin of time of our theoretical $V$ light curve with
the observation, so that the early rising phase of our model matches well 
the rising phase in the Evryscope data \citep{qui24}.
Thus, we adopt the origin
of time to be $t({\rm B})= 0 = t_{\rm OB}=$JD 2,459,377.68 in our models,
where $t({\rm B})$ is the time at stage B in Figure \ref{hr},
that is, the outburst day of $t_{\rm OB}$.  This time is 0.0103 days
before the reference time ($=t_0$) adopted by \citet{sok23}.

Both of the models A and B catch the main properties of light curve; 
a steep rise and gradual decline. The good agreement of 
overall timescale indicates that our choice of 1.35 $M_\sun$ 
is appropriate. Model A (black line) shows a better agreement 
around the optical peak 
because a larger nuclear burning rate $L_{\rm nuc}^{\rm max}$ 
results in faster expansion of the envelope. 
Also a larger ignition mass results in a more extended 
envelope at the optical peak and larger wind mass loss rates.



\begin{figure*}
\epsscale{0.85}
\plotone{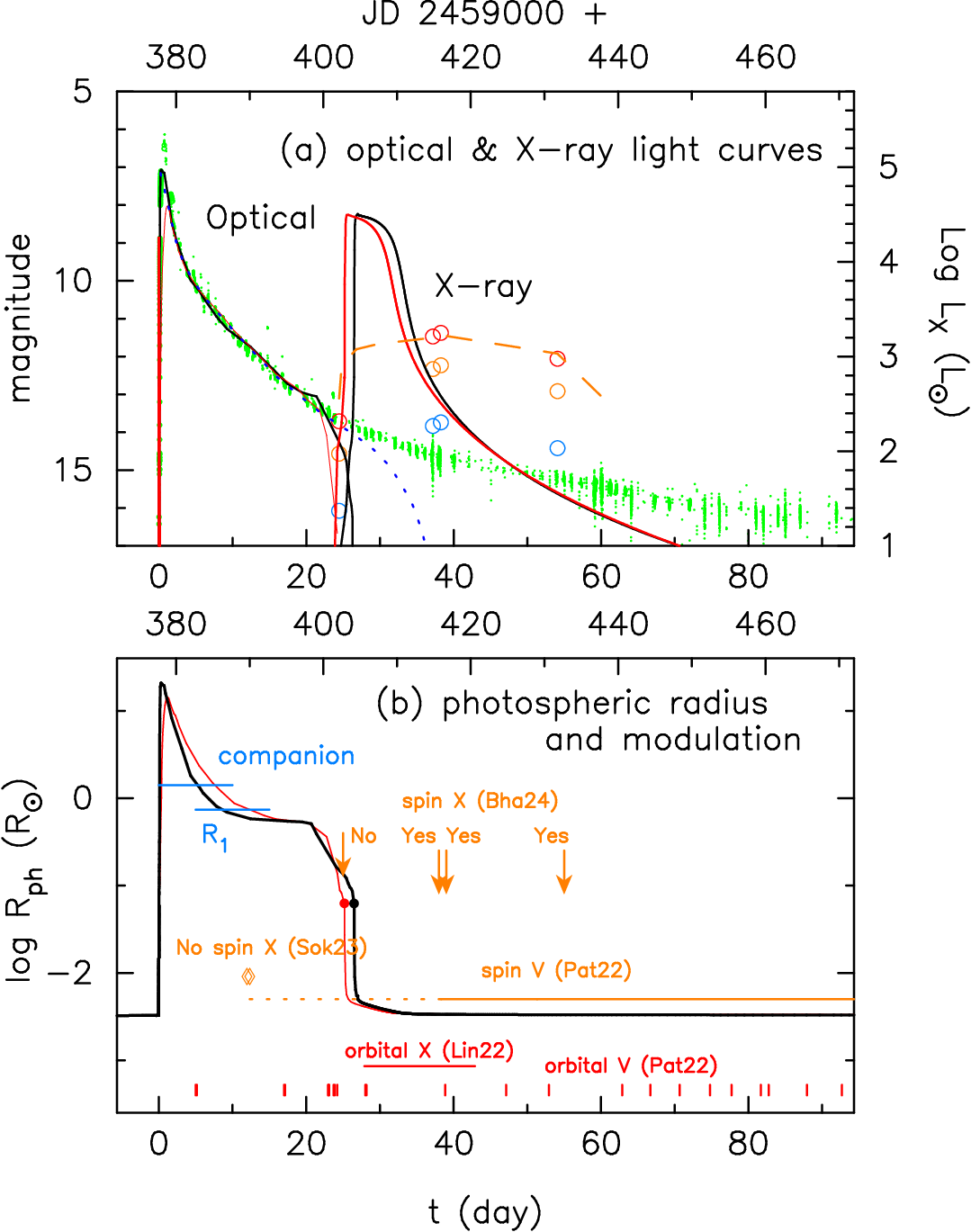}
\caption{
(a) Theoretical optical $V$ and X-ray (0.3--1.0 keV) light curves
of our $1.35~M_\sun$ models as well as the observational data.
The black lines are for model A ($\dot{M}_{\rm acc}= 1\times 10^{-11} ~M_\sun$
yr$^{-1}$) and the red lines for model B ($\dot{M}_{\rm acc}= 
5\times 10^{-10} ~M_\sun$ yr$^{-1}$).
The lower abscissa show the origin of time for our theoretical light curves
and the upper abscissa corresponds to JD date for observation.
The red open circles connected by the orange dashed line
indicate the X-ray luminosities calculated from the fluxes obtained by 
\citet{bha24} with our distance of 8.9 kpc.
For comparison, we added the orange and blue 
open circles for the distance of $d=6$ and 3 kpc, respectively.
The blue dotted line indicates the free-free light curve
calculated from the steady state sequence described
in Section \ref{sslightcurve}).
(b) Temporal change of the photospheric radius of the WD. 
The filled circles on the line correspond to stage I when the 
optically thick winds stop. 
The two blue lines indicate the orbital separation $a$ of the companion
and the effective Roche lobe radius $R_1$ of the WD. 
Orange open diamonds: no spin detection in X-ray \citep{sok23}.
Horizontal orange line: the spin period is detected in the $V$ 
light curve \citep{pat22}. 
Downward orange arrows: whether the spin period is detected or not 
with X-ray \citep{bha24}.
Horizontal red line: the orbital period is detected in X-rays \citep{lin22}.
Short vertical red lines: the epochs of the $V$ minimum 
in the orbital modulation \citep{pat22}.
}\label{TRV.compari}
\end{figure*}

The downward arrow indicates the epoch when the photospheric 
radius shrinks and the companion star 
appears from the extended envelope.   
The temporal change of the WD photospheric radius will be shown later in 
Figure \ref{TRV.compari}.
As before, we assume the companion mass of $M_{\rm comp}=0.26~M_\sun$ 
after \citet{qui24}. Then the binary separation is $a=1.41~R_\sun$ 
for $M_{\rm WD}=1.35 ~M_\sun$ 
with the orbital period of $P_{\rm orb}=0.1529$ days ( =3.67 hr).  
There is no indication in the light curve 
of extra mass ejection (or extra brightness) related to the companion 
motion. 
%
%
We conclude that the decay phase light curve of V1674 Her 
is well explained by free-free emission from ejecta 
of which the mass loss rate is owing to the optically thick winds,    
i.e., continuum-radiation-driven mass loss that occurs 
deep inside the photosphere \citep{fri66,kat94h}.  


\subsection{X-ray Light Curve}

X-rays are detected with various satellites, with the X-ray spectra of
the nuclear spectroscopic telescope array \citep[NuSTAR: ][]{sok23},
Swift/X-Ray Telescope (XRT) and Chandra \citep{dra21}, 
Neutron star Interior Composition Explorer
\citep[NICER : ][]{lin22, ori22}, and AstroSat \citep{bha24}. 
The X-ray count rate raised on day $\sim 20$ and 
stayed high following 40 days. 
The spectra are dominated by soft component ($< 1$ keV). 
The count rate is highly modulated 
with a period of 501 s, 
and such a modulation is thought to be 
an indication of emission originated from the vicinity of magnetic poles
rather than the entire WD surface \citep{dra21, lin22, ori22, bha24}. 



Another possible interpretation of the strong modulation is 
occultation of the WD surface. 
Hydrogen burning occurs entire WD surface which 
is mostly occulted by dense stream, funneling into a magnetic pole. 
The soft X-ray spectrum ($< 1$ keV) seems to support this idea of 
occultation of hydrogen-burning WD, 
rather than magnetic pole emission.

Figure \ref{TRV.compari}(a) shows observed X-ray luminosity at four
epochs calculated from the fluxes in Table 1 of \citet{bha24}.
These absorbed fluxes are plotted in Figure \ref{TRV.compari}a. 
Here we assumed the three distances to V1674 Her:
$d=8.9$ kpc (red circles), $d=6$ kpc (orange circles), 
and $d=3$ kpc (blue circles), In Section \ref{sec_distance},
we determine the distance to V1674 Her to be 8.9 kpc. 
Note that these fluxes are orbital mean  
and therefore the intrinsic flux could be much higher. 


For a simple comparison, 
we have connected the four data obtained by \citet{bha24}  
by the dashed line. 
This line is based on our eye impression and not accurate, 
but to demonstrate that the count rates are relatively 
flat before day 40 and no sharp peak between 
day 20 and 40 \citep[see also, e.g., Figure 1 of ][]{bha24}. 

The X-ray rising time (X-ray turn-on time) is very consistent with 
the first data point of \citet{bha24}, but our theoretical flux lasts 
only $\sim 10$ days and decays much faster than was observed. 
This discrepancy will be discussed in Section \ref{acc_rate_postnova}.



\subsection{WD photospheric radius and periodic modulation}\label{sec_radius}

\subsubsection{Temporal change of WD radius}

After the onset of thermonuclear runaway, the photospheric radius of 
hydrogen-rich envelope of the WD quickly increases to reach 
maximum expansion where the wind mass loss rate also reaches maximum.
In the decay phase, the photospheric radius gradually 
shrinks as the envelope mass decreases with time due to wind mass loss. 
Such a temporal change of the photospheric radius is 
plotted in Figure \ref{TRV.compari}b. 
The optically thick winds stop at the filled circle in each model  
in Figure \ref{TRV.compari}b  (stage I in Figure \ref{hr}).
Around this epoch, the photosphere rapidly shrinks and finally returns 
to the value before the outburst.


In the extended phase, the photospheric radius largely expands to
$R_{\rm ph}=21~R_\sun$ for model A and $R_{\rm ph}=14~R_\sun$ for 
model B. In both cases the envelope expands 
far beyond the effective Roche lobe radius of the WD 
($R_1=0.74~R_\sun$), and the orbit of the companion ($a=1.4~R_\sun$). 
If the envelope expands spherically,   
the binary is completely hidden below the photosphere 
and no orbital/spin modulations would be detectable.


After the winds stop, the WD photosphere
rapidly shrinks to the size as was in the pre-explosion stages. 
Then, the companion star, accretion disk and magnetic poles 
would appear and the optical and X-ray modulations 
associated with the orbital motion and spin of the WD 
would be visible. 
In what follows, we examine these modulations one by one based on
our 1.35 $M_\sun$ WD model.

\subsubsection{Orbital modulation of optical magnitude}

The optical light curve shows orbital period variation.  
\citet{pat22} listed measured times of the minimum light in the 
$V$ light curve, of which the epochs are indicated in 
Figure \ref{TRV.compari}(b) with the vertical short red lines.  
The orbital period variation has been detected from 
the early expanded stage until the later phase.  
 
\citet{pat22} also presented the average waveform 
for one orbital cycle of the $V$ light curve of $t > 31$ days. 
This waveform shows two clear asymmetric dips. 
One is an asymmetric deep dip at phase 0.0, that 
occurs when the companion hides the WD. 
The another is a wide shallow asymmetric dip 
at phase 0.5--0.7. 

Such an asymmetric light curve has been discussed 
for a high inclination binary 
in which an irradiated disk is 
partly elevated by the L$_1$ accretion stream 
\citep[e.g.,][]{buc89,sch97mm}.
Comparing with \citet{sch97mm}'s light curve model 
for the LMC SSS Cal 87, 
we see that the wide shallow dip at phase 0.5--0.7 
could appear when the irradiated side of the elevated disk faces 
opposite direction to us. 
In other words, the asymmetric orbital variation of V1674 Her 
strongly suggests the presence of survived/reformed 
accretion disk at $t > 31$ d and a hydrogen burning WD. 

The first two adjacent red lines (on day 5.0 and 5.1) in  
Figure \ref{TRV.compari}b are located at a 
very early stage, i.e.,  the companion star should be still 
embedded in the extended envelope.
\citet{pat22} noted that these early timings are not 
necessarily continuous with the family of later timings.  
Moreover, the optical variation in this early stage  
shows a different shape from those 
in the later stages \citep[see Figure 1 of ][]{pat22}. 
Thus, we may conclude that this early time variation may
reflect some asymmetric structure of the 
expanding ejecta but may not indicate the irradiated asymmetric disk. 
The temporal evolution of the waveform is 
consistent with the photospheric radii of our evolution model. 



\subsubsection{Orbital modulation in X-ray}

After the optically thick winds stop, 
the magnetic poles could emerge from the photosphere. 
\citet{lin22} presented X-ray orbital modulation,  
subtracting pulsed component from NICER data 
obtained from 2021 July 10 to 25 ($t=28$--43 day). 
%
The X-ray light curve shows double-humped structure 
that the authors attributed to the occultation 
by the companion or the disk. 
The dip at phase 0.0 is slight asymmetric, indicating 
the occultation both by the companion star and disk 
as in the optical orbital modulation. 

The second peak (phase $> 0.7$) is much lower than the first peak, 
indicating stronger effects by the elevated disk. 
In this way, the X-ray orbital modulation can be explained 
with a similar disk structure as suggested in the optical variation.


This X-ray orbital modulation is detected in the data between 
$t= 28$--43 day as indicated with the horizontal red line 
with ``orbital X (Lin22)'' in Figure \ref{TRV.compari}b. 
This interval is consistent with the small radii of our models
A and B.

\subsubsection{Spin period modulation}

\citet{sok23} reported non-detection of spin period modulation 
in the Swift/XRT data on day 11.9 and 12.3. These two epochs 
are shown by orange diamonds in Figure \ref{TRV.compari}b. 
In this stage, the WD photosphere is still as large as comparable 
to its Roche lobe size, $R_1$.  The X-ray emitting region 
(magnetic poles or WD surface) could
be still embedded in the WD photosphere.  
Non detection of  spin modulation is consistent with 
our models A and B. 

\citet{bha24} detected X-ray spin-period on day 37, 38, and 54.
In an earlier day, on day 24, the count rate is low 
(one tenth of those on day 37 and 38) 
and no clear orbital periodicity is detected.  
These non-detection/detection signals are indicated by No/Yes words 
together with downward orange arrows in Figure \ref{TRV.compari}b. 
These no/yes detections of X-ray spin signal 
are consistent with our nova models. 

\citet{pat22} reported that rapid periodic signals in optical band
first appeared around 0.01 mag full amplitude near day 12 
and steadily grew, reaching $\sim 0.09$ mag on day 350. 
If these small amplitude variations are really connected
to the spin period, it would reflect an asymmetric structure of ejecta,  
which is valuable information for mass ejection in
intermediate polars.

\begin{figure*}
\epsscale{0.75}
\plotone{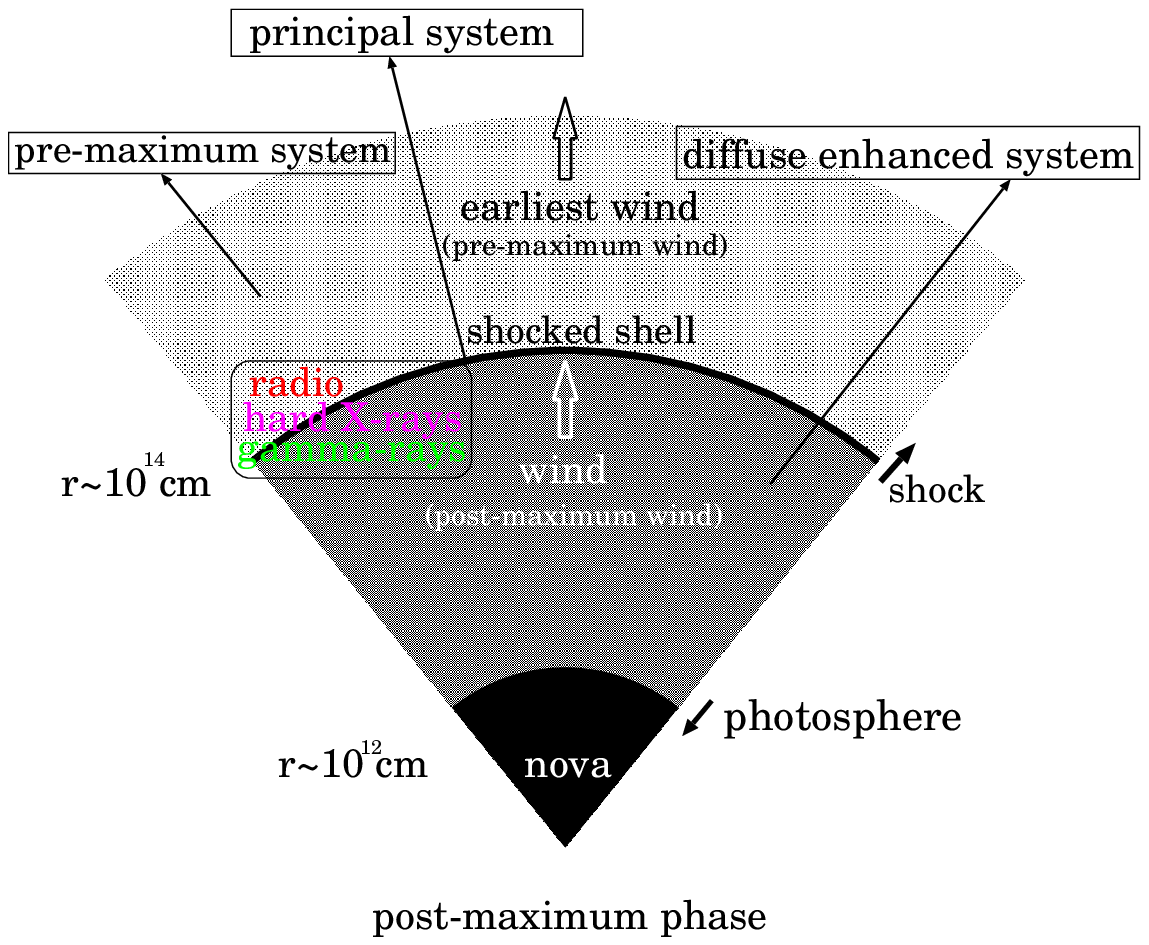}
\caption{
Schematic illustration of ejecta configuration of V1674 Her
in the post-maximum phase.
A shock wave arose just after the maximum expansion of the photosphere
and has already moved far outside the WD photosphere (and the binary).
The three emission/absorption line systems raised by \citet{mcl42}, i.e.,
pre-maximum, principal, and diffuse enhanced, originate
from the pre-maximum wind (earliest wind), shocked shell, and
post-maximum wind (inner wind), respectively.
The shocked shell emits most part of 
radio, hard X-rays, and gamma-rays.
This figure is taken from Figure 2(c)
of \citet{hac23k} with a modification.
We assume that ejecta are spherically symmetric.
}
\label{v1674_her_shock_configuration}
\end{figure*}

\subsection{X-ray Temperature in the SSS Phase}

\citet{dra21} obtained blackbody fit temperatures from  
the Swift/XRT data, that is, 
$kT \sim 54$ eV between day 18.9 and day 27.7, 
$kT \sim 130$ eV on day 39.6, 
$\sim 100$ eV by day 74.2. 
\citet{bha24} estimated the blackbody temperature to be 
$\sim 110$ eV at the four epochs indicated by the four arrows 
denoted as ``spin X (Bha24)'' in Figure \ref{TRV.compari}b.  
These values are  much lower and incompatible to those 
expected from an intermediate polar 
\citep[see, e.g., discussion of ][]{sok23}. 

These temperatures are broadly consistent with 
our theoretical values of $\log T_{\rm ph}^{\rm max}({\rm K})
\sim 6.1$  =109 eV if these X-rays originate from a hydrogen burning WD
of $1.35 ~M_\sun$.

\subsection{Hard X-ray temperature and luminosity}
\label{hard_x-ray_shock}

\citet{sok23} reported hard X-ray emission in V1674 Her 
observed on day 11 with NuSTAR and Swift, of which
the hard X-rays originate from 
optically thin thermal plasma that was shock-heated
to $k T_{\rm shock} = 4$ keV.
Here we will explain the hard X-ray properties in V1674 Her  
based on our theoretical results.

\subsubsection{shock formation and McLaughlin's emission/absorption
line systems}  

\citet{hac22k, hac23k} proposed a shock formation mechanism 
based on the nova model calculated by \citet{kat22sha}. This  
fully self-consistent calculation shows 
a temporal change of the ejecta velocity during the outburst. 
The velocity at the photosphere decreases with time 
before maximum expansion of the photosphere, but turns
to increase after that.  Such a velocity evolution has been  
observationally confirmed in several classical novae \citep{ayd20ci}.
In the post-maximum phase, the wind ejected later is catching up
the matter previously ejected, which causes a strong shock
outside the photosphere \citep{hac22k}.

In our $1.35 ~M_\sun$ WD model, the photospheric velocity (the right edge
of each line) also decreases with time toward stage G (the maximum expansion
of the photosphere) as in Figure \ref{rhov.m135}(b),
but increases after that as in Figure \ref{rhov.m135}(d).
The winds are all supersonic near the photosphere because the velocity
becomes larger than the isothermal sound velocity outside the critical point, 
which is located deep inside the photosphere. Thus, matter will be compressed,
which causes a strong shock wave (reverse shock).
Therefore, the internal shock should arise just after 
the maximum expansion of the photosphere.
The $V$ band luminosity calculated with Equation 
(\ref{free-free_flux_v-band})
reaches maximum at the maximum expansion of the photosphere.
Thus, a shock should arise just after the optical $V$ maximum.

Figure \ref{v1674_her_shock_configuration} illustrates such 
a configuration that a shocked shell is formed outside 
the photosphere and moving outward.  The mass of the shocked shell
($M_{\rm shell}$) increases with time, 
and finally reaches more than $> 90$\% of the total 
ejecta mass \citep{hac22k}, because the nova wind continuously blows
and its mass is added continuously to the shocked shell.

From the observational point of view,
\citet{mcl42} proposed three emission/absorption line systems in
nova spectra before and after optical maximum of novae; pre-maximum, principal,
and diffuse enhanced emission/absorption line systems.  McLaughlin described
their features, ``The pre-maximum absorption lasts from the earliest date
at which any nova has been observed on the rise until at least a day or
a few days after maximum light.  The principal absorption spectrum becomes
dominant within the first few days after maximum. Its displacement is
greater than that of the pre-maximum spectrum.  The diffuse enhanced
absorption is a set of very strong and diffuse lines of hydrogen and
ionized metals, with a displacement roughly double that of the principal
spectrum.''
Although the origins of these absorption line systems have long been
debated \citep[e.g.,][]{wil92},
\citet{hac22k} interpreted that the principal absorption/emission 
line system originates from the shocked shell, and the diffuse enhanced
absorption/emission line system is from the inner wind after the optical
maximum, as illustrated in Figure \ref{v1674_her_shock_configuration}.
The coincidence of two epochs, shock formation and maximum brightness,
as mentioned above, can reasonably explain the appearance of the principal
and diffuse enhanced absorption/emission line systems as well as the
disappearance of the pre-maximum absorption/emission line system.  

In V1674 Her, 
\citet{ayd21sc} reported a P-Cygni profile at $\sim 3000$
km s$^{-1}$ on day 2.67.
We regard that this velocity is the principal system of
$v_{\rm p}$ and identified to be 
$v_{\rm shock}= v_{\rm p}\approx 3000$ km s$^{-1}$.
Slightly later, a high velocity component of $\sim 5000$ km s$^{-1}$
troughs of absorption appeared.
This higher velocity component is regarded as the diffuse enhanced system
of $v_{\rm d}$ and identified to be
$v_{\rm wind}= v_{\rm d}\approx 5000$ km s$^{-1}$. 
Here, $v_{\rm p}$ and $v_{\rm d}$ are the velocities of the principal and
diffuse enhanced systems, respectively, and
$v_{\rm wind}$ and $v_{\rm shock}$ are the velocities of the inner wind
and shock, respectively.

\subsubsection{temperature and luminosity of shock}
\label{shock_temperature_luminosity}

The temperature of shocked matter and shock energy can be estimated from
the difference between the ejecta (inner wind) velocity and speed of shock.
The temperature just behind the shock (reverse shock) is estimated to be
\begin{eqnarray}
kT_{\rm sh}& \sim & {3 \over 16} \mu m_p
\left( v_{\rm wind} - v_{\rm shock} \right)^2 \cr
& \approx & 1.0 {\rm ~keV~}
\left( {{v_{\rm wind} - v_{\rm shock}} \over
{1000 {\rm ~km~s}^{-1}}} \right)^2,
\label{shock_kev_energy}
\end{eqnarray}
where $k$ is the Boltzmann constant,
$T_{\rm sh}$ is the temperature just after the shock
\citep[see, e.g.,][]{met14hv},
$\mu$ is the mean molecular weight ($\mu =0.5$ for hydrogen plasma),
and $m_p$ is the proton mass.
Substituting $v_{\rm shock}= v_{\rm p}=3000$ km s$^{-1}$ and
$v_{\rm wind}= v_{\rm d}=5000$ km s$^{-1}$,
we obtain the post-shock temperature
$k T_{\rm sh}\sim 4$ keV.
This shock temperature is consistent with the 
NuSTAR and Swift observations on day 11, that is,
$k T_{\rm shock}\approx 4$ keV \citep{sok23}. 

\citet{sok23} interpreted that the temperature of shock heated plasma
is directly connected to the kinetic energy loss at the shock wave,
and obtained the shock velocity of $v_{\rm shock}=1700$ km s$^{-1}$
from their Equation (3).  However, no strong emission/absorption lines
of $1700$ km s$^{-1}$ have ever been reported.  On the other hand,
our interpretation of shock velocity and temperature is compatible
with the observation.
Moreover, our shock formation mechanism naturally explains the origin
of McLaughlin's absorption/emission line systems of novae \citep{hac22k}.

Mechanical energy of the wind is converted to thermal energy
by the reverse shock \citep{met14hv} as
\begin{eqnarray}
L_{\rm sh}& \sim & {{9}\over {32}} {\dot M}_{\rm wind}
{{( v_{\rm wind} - v_{\rm shock} )^3} \over {v_{\rm wind}}} \cr
&=& 1.8\times 10^{37}{\rm ~erg~s}^{-1}
\left( {{{\dot M}_{\rm wind}} \over
{10^{-4} ~M_\sun {\rm ~yr}^{-1}}} \right) \cr
 &  & \times
\left( {{{v_{\rm wind} - v_{\rm shock}} \over {1000{\rm ~km~s}^{-1}}}}
\right)^3
\left( {{{1000{\rm ~km~s}^{-1}} \over {v_{\rm wind}}} }\right).
\label{shocked_energy_generation}
\end{eqnarray}
Substituting $\dot{M}_{\rm wind}= 1.2 \times 10^{-4} ~M_\sun$ yr$^{-1}$
from our model A ($1.35 ~M_\sun$ WD with $\dot{M}_{\rm acc}=1\times
10^{-11} ~M_\sun$ yr$^{-1}$),
$v_{\rm shock}= v_{\rm p}=3000$ km s$^{-1}$, and
$v_{\rm wind}= v_{\rm d}=5000$ km s$^{-1}$, both from the observation
\citep{ayd21sc},
we obtain the post-shock energy flux of
$L_{\rm sh} \sim 3.5\times 10^{37}$ erg s$^{-1}$ just after maximum.
This energy flux decreases to 
$L_{\rm sh} \sim 6\times 10^{36}$ erg s$^{-1}$ on day 11.
\citet{sok23} obtained the energy flux between 0.3--78 keV to be
$L_{\rm X}= 1.4\times 10^{34}$ erg s$^{-1}$
 $(8.9 {\rm ~kpc}/ 6.3{\rm ~kpc})^2 =
2.8\times 10^{34}$ erg s$^{-1}$ on day 11,
which is about 0.5\% of the $L_{\rm sh}$.
Here we corrected the flux from  \citet{sok23}'s distance 
of  6.3 kpc to our distance of $d= 8.9$ kpc (see Section \ref{sec_distance}). 

\subsubsection{column density of shocked shell}
The column density of hydrogen is estimated from
$M_{\rm shell}= 4 \pi R_{\rm sh}^2 \rho h_{\rm shell}$,
where $M_{\rm shell}$, $\rho$, and $h_{\rm shell}$ are 
the mass, density, and the thickness of the shocked shell.
If we take an averaged velocity of shell $v_{\rm sh}=
v_{\rm shell}= v_{\rm shock}= 3000$ km s$^{-1}$,
the shock radius is calculated from $R_{\rm sh}(t)= v_{\rm shock}\times t$.
This reads
\begin{eqnarray}
N_{\rm H} & = & {{X \over m_p} {{ M_{\rm shell} }
\over {4 \pi R^2_{\rm sh}}}} \cr
 & \approx & 4.8\times 10^{21} {\rm ~cm}^{-2}
\left({X \over {0.5}}\right)
\left( {{M_{\rm shell}} \over {10^{-6} M_\sun}} \right)
\left( {{R_{\rm sh}} \over {10^{14} {\rm ~cm}}} \right)^{-2}
\cr
 & \approx & 6.4 \times 10^{21} {\rm ~cm}^{-2}
\left({X \over {0.5}}\right)
\left( {{M_{\rm shell}} \over {10^{-6} M_\sun}} \right) \cr
& & \times
\left( {{v_{\rm shell}} \over {1000 {\rm ~km~s}^{-1}}} \right)^{-2}
\left( {{t} \over {10~{\rm day}}} \right)^{-2}.
\label{column_density_hydrogen_time}
\end{eqnarray}
Thus, the column density decreases from 
$N_{\rm H} \approx 9\times 10^{21}$ cm$^{-2}$ on day $t=3$ to 
$\approx 0.6\times 10^{21}$ cm$^{-2}$ on day $t=11$ for 
the shell mass of $M_{\rm shell}=1 \times 10^{-6}~M_\odot$ 
in model A.

\citet{sok23} concluded that  
the NuSTAR spectrum on day 11 is consistent with 
the Galactic absorption $N_{\rm H}\sim
4\times 10^{21}$ cm$^{-2}$ with 
no intrinsic absorbing column. 
The neutral hydrogen column density toward V1674 Her is slightly
smaller, $N_{\rm H}\sim 3\times 10^{21}$ cm$^{-2}$
\citep{hi4pi16}, than this value.
Therefore, our column density of the shocked shell,
$N_{\rm H}\approx 0.6\times 10^{21}$ cm$^{-2}$, on day $t=11$ is consistent
with \citet{sok23}'s estimate. 


\citet{sok23} suggested a 10 ks ($=10^4$ s) timescale of the NuSTAR X-ray 
flux variability.  This timescale requires that the radius of the
shocked shell (or the size of a shock front)
is as large as $c t_{\rm var}= (3\times 10^{10}$ cm s$^{-1})\times (10^4$ s)
$=3\times 10^{14}$ cm on day $t=11$, where $c$ is the speed of light and
$t_{\rm var}$ the timescale of variability.  The shocked shell of our model
is located at $R_{\rm sh}= v_{\rm shell} \times t = 3000$ km s$^{-1}
\times $11 days $\approx 3\times 10^{14}$ cm on day $t=11$,
being consistent with the timescale of the X-ray flux variability.

\subsubsection{How long is the shock alive?}

\citet{hac23k} estimated the shock duration $\tau_{\rm shock}$ by
\begin{equation}
\tau_{\rm shock}= {{t_{\rm ws}} \over
{\left( 1- {{v_{\rm p}} \over {v_{\rm d}}}\right)}},
\label{duration_of_shock}
\end{equation}
where $t_{\rm ws}$ is the wind stopping time.
Substituting $v_{\rm sh}\approx v_{\rm p}\sim 3000$ km s$^{-1}$
(principal system), $v_{\rm ph}\approx v_{\rm d}\sim 5000$ km s$^{-1}$
(diffuse enhanced system), and $t_{\rm ws}=26$ days (the wind duration
since the shock arises) into Equation (\ref{duration_of_shock}),
we obtain the shock duration of $\tau_{\rm shock}= 26/0.4= 65$ days.
Therefore, in V1674 Her, we expect hard X-ray emission until day 65.  
This time correspond to the end of the SSS phase (as later shown in
Figure \ref{v1674_her_kt_eri_v339_del_v_x_stretching}, 
or, e.g., Figure 1 of \citet{dra21}).

\subsection{Radio flux}
\label{radio_flux_shock}

\citet{sok23} also presented the radio flux and brightness temperature
at 2.6--34.0 GHz 
from day 3.2 to day 409.1.  The flux $F_\nu$ approximately increases
with time, along the $F_\nu \propto t^2$ law,
where $t$ is the time from the outburst.
The fluxes reached maximum on day $t=44.1$ and turned to decay 
on day $t=74.0$.  
The brightness temperature of radio also reached maximum on day $t=44.1$
and started to decay on day $t=74.0$ \citep[see Figure 7 of ][]{sok23}.

This rising trend could be explained as the increasing
shock surface area of $4\pi R_{\rm shock}^2 \propto (v_{\rm shock} t)^2$
if the radio emission is optically thick.
The epoch of turn (from rise to decay) is roughly consistent 
with the end day of the shock, 
day 65 obtained from Equation (\ref{duration_of_shock}).
After that day, a rarefaction wave penetrates the dense shell, and
the density of the shell decreases.
Therefore, the radio emission becomes optically thin and 
the flux starts to decay.

\subsection{GeV Gamma-ray Light Curve}

0.1--300 giga electron volt (GeV) gamma-rays were detected
with the Fermi/Large Area Telescope 
\citep[Fermi/LAT, see][for details]{sok23}. 
Sokolovsky et al. obtained a light curve with 6 hr bins in their
Figure 1 during $t_0\pm 2$ days, where $t_0$ is their origin of time,
$t_0=$JD 2,459,377.6903.  The first detection of gamma-rays 
is on their day $0.38\pm 0.25$ ($=$ on our day $0.39\pm 0.25$).


In our model, 
the internal shock should arise just after the maximum expansion
of the photosphere (stage G). 
The time of epoch G in model A, $t=0.32$ day, 
is very close to the emergence
time of gamma-rays, $0.39\pm 0.25$ day obtained by \citet{sok23}.

In the shock formation mechanism proposed by \citet{hac22shock}, 
gamma-rays appear only after theoretical optical maximum. 
In V1674 Her, however, there are only a few visual data 
between day 0.25 and 0.65, during which the light curve seems to be flat
\citep[see Figure 1 of ][]{sok23}. It is not clear 
whether our theoretical $V$ maximum coincides with observed 
$V$ maximum.  
%

The post-shock energy flux is estimated to be
$L_{\rm sh} \sim 3.5\times 10^{37}$ erg s$^{-1}$ just after maximum
for our model A.  
\citet{sok23} obtained the gamma-ray energy flux
of $L_\gamma = 3.2\times 10^{36}$ erg s$^{-1} 
~(8.9 {\rm ~kpc}/6.3 {\rm ~kpc})^2 \approx 6 \times 10^{36}$ erg s$^{-1}$
between 0.1--300 GeV.  This gamma-ray flux
is 17\% of the post-shock energy flux $L_{\rm sh}$ near
just after maximum expansion of the photosphere (stage G).




\section{Discussion}\label{sec_discussion}

\subsection{Distance and Reddening} \label{sec_distance}
 
\begin{figure}
\epsscale{1.15}
\plotone{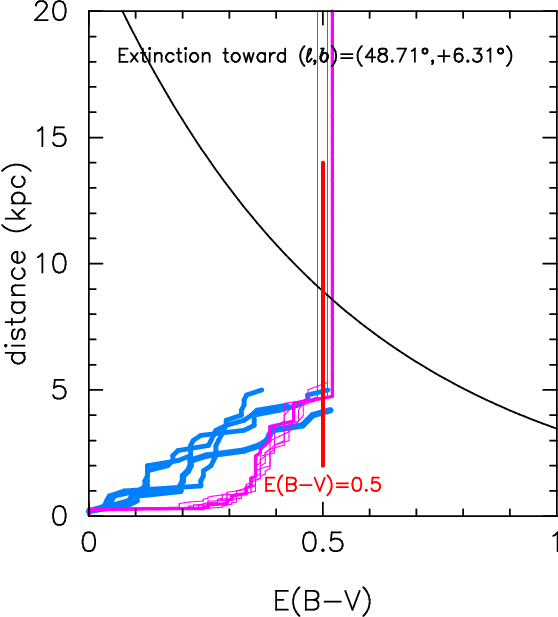}
\caption{
The distance-reddening relations toward V1674 Her whose galactic coordinates
are $(\ell, b)= (48\fdg71, +6\fdg31)$.
The black line denotes the relation of Equation (\ref{distance_reddening_law})
together with $(m-M)_V=16.3$ for V1674 Her.  The thin magenta lines are
the sample distance-reddening relations given by \citet{gre19}
while the thick magenta line is their best-fit line for them.
Here, we use the relation of $E(B-V)= 0.884\times$(Bayestar19) \citep{gre19}.
We add four distance-reddening relations
(thick cyan-blue lines) of \citet{chen19}, which correspond to four
nearby directions toward V1674 Her, i.e., the galactic coordinates of
$(\ell, b)= (48\fdg65, +6\fdg25)$, $(48\fdg65, +6\fdg35)$,
$(48\fdg75, +6\fdg25)$, and $(48\fdg75, +6\fdg35)$.
The two relations (black and magenta lines) cross at the distance of
$d\approx 8.9$ kpc and $E(B-V)\approx 0.5$.
}\label{distance_reddening_v1674_her}
\end{figure}

The distance to V1674 Her is not well constrained. 
Various authors have presented various values from $\sim 2$ to $\sim 6$ kpc 
\citep{bai21gaia, woo21, qui24, schaefer22, sok23, hab24}.  
Negative Gaia parallax indicates a rather long distance. 

Here we estimate the distance to V1674 Her 
using the $V$ band distance modulus 
$(m-M)_V= 16.3\pm 0.2$ obtained in Appendix \ref{time_stretching_method}. 
Figure \ref{distance_reddening_v1674_her} shows the distance-reddening 
relation (black line)
\begin{equation}
(m-M)_V= 5 \log (d / {\rm 10~pc}) + 3.1 E(B-V),
\label{distance_reddening_law}
\end{equation}
where we adopt $(m-M)_V= 16.3$ and $R_V= 3.1$ \citep{rie85}.


This figure also shows two 3D extinction maps taken from \citet{gre19} and
\citet{chen19}.  
Note that we use the relation of $E(B-V)= 0.884\times$(Bayestar19)
as suggested by the website\footnote{http://argonaut.skymaps.info} of
Baystar19 3D map \citep{gre19}. 
From the crossing point of the black line of Equation
(\ref{distance_reddening_law}) and Green et al.'s magenta lines
we obtain $d=8.9\pm 1$ kpc and $E(B-V)= 0.5 \pm 0.05$.
The reddening of $E(B-V)= 0.5$ is supported by \citet{schlaf11f}'s 2D
reddening map of $E(B-V)= 0.4985\pm 0.0191$.
In the present paper, we adopt $(m-M)_V= 16.3$, $d=8.9$ kpc, 
and $E(B-V)= 0.5$ unless otherwise specified.

\subsection{Chemical Composition of the Envelope}
\label{sec_composition}

We have assumed the solar composition, $X=0.7$, $Y=0.28$, and $Z=0.02$
for the accreted matter.
To mimic heavy element enrichment, often observed in nova ejecta,
we have simply increased carbon mass fraction by 0.1
and decreased helium fraction by the same amount at the
onset of thermonuclear runaway \citep{kat22shapjl,kat22shc}.
After the shell flash sets in, convection widely spread throughout
the envelope and nuclear products are mixed into the upper layer.
Then, the chemical composition becomes almost uniform throughout 
the envelope.  See \citet{kat22sha} for details of our treatment of convection 
and element mixing.  At the optical peak, the chemical composition of the
envelope changes to $X=0.55$, $Y=0.29$, $X_{\rm N}=0.14$, and $Z=0.02$
and keeps this composition after that, where $X_{\rm N}=0.14$ is extra
nitrogen by mass.  Therefore, the ejecta is nitrogen-rich.


\citet{sok23} suggested a non-solar, heavy element enrichment
in the ejecta from the NuSTAR X-ray spectra.  
While, modeling the observed
optical spectra with the code CLOUDY, \citet{hab24} obtained
the He, O, N, Ne, and Fe fractions to be overabundant relative
to the solar values.
If we assume that the other elements are solar in abundance,
their results would be converted to $X=$0.41--0.45, $Y=$0.43--0.5,
and $Z=$0.08--0.11 by mass.

These results are in qualitative agreement with our model,
in the sense of hydrogen deficient and heavy element enrichment.
The dependence of our model light curve on the chemical composition
will be examined in Section \ref{sec_steadystate}.

\begin{figure*}
\gridline{\fig{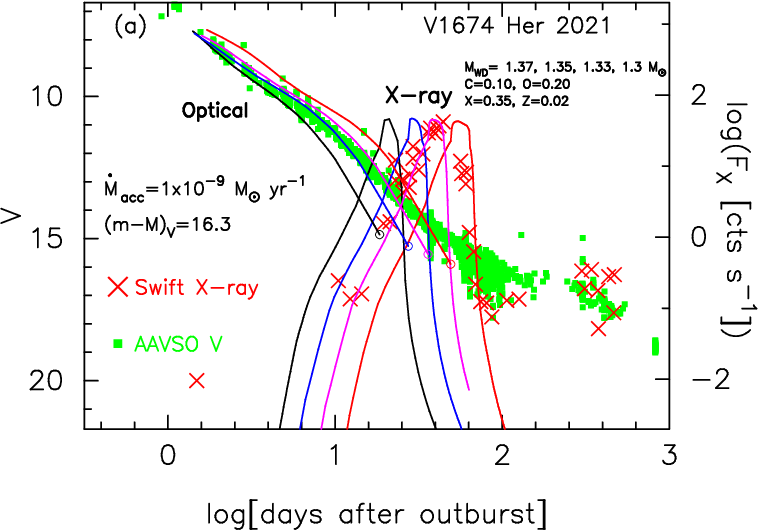}{0.45\textwidth}{}
          \fig{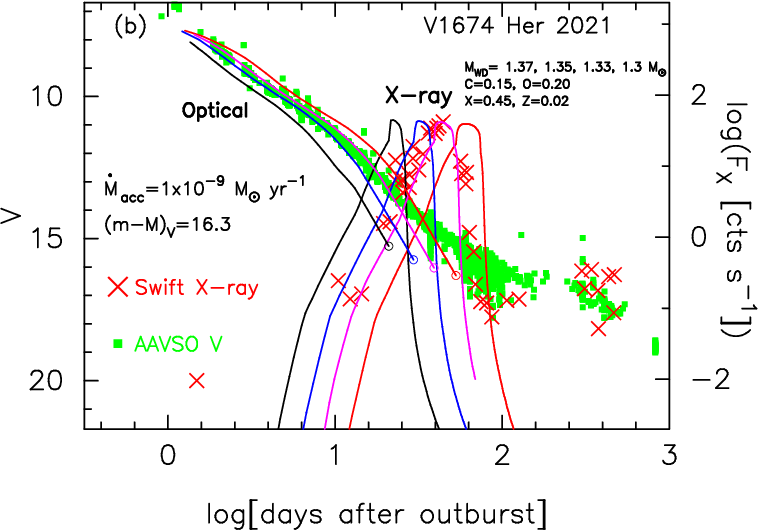}{0.45\textwidth}{}
          }
\gridline{
          \fig{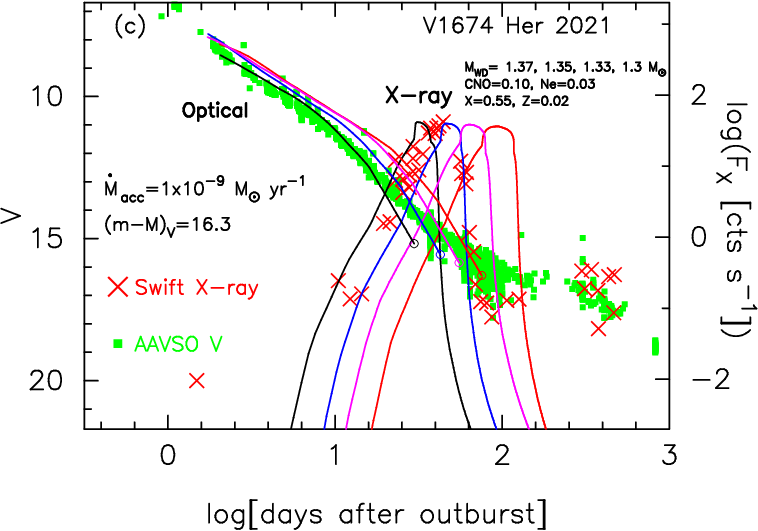}{0.45\textwidth}{}
          \fig{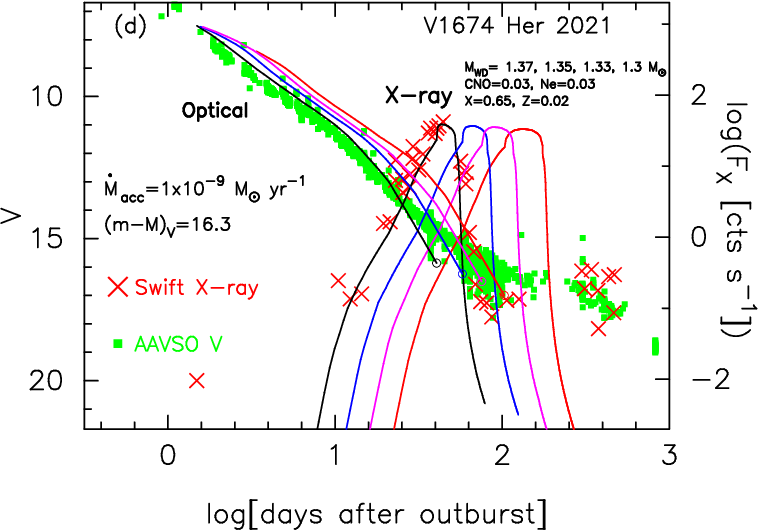}{0.45\textwidth}{}
          }
\caption{
The model $V$ light curves, for the
distance modulus in the $V$ band of $(m-M)_V=16.3$, and
X-ray (0.3-10.0 keV) light curves,
for different sets of WD mass and chemical composition.  The all $V$
and X-ray model light curve data are taken from \citet{hac25kv392per}.
The mass-accretion rate onto the WD is fixed to be 
$\dot{M}_{\rm acc}= 1\times 10^{-9} M_\sun$ yr$^{-1}$ for all models.
The open circle at the end of each line indicates the epoch when 
the optically thick winds stops.
We add the observed $V$ magnitudes and X-ray count rates; 
the $V$ data are taken from
AAVSO and the X-ray data are from the Swift website \citep{eva09}. 
The outburst day is assumed to be $t_{\rm OB}=$JD 2,459,377.68.
(a) The chemical composition of the envelope is CO nova 2 (CO2):
$1.3~M_\sun$ (red), $1.33~M_\sun$ (magenta),
$1.35~M_\sun$ (blue), and $1.37~M_\sun$ (black) WDs.
(b) CO nova 3 (CO3): 
 $1.3~M_\sun$ (red), $1.33~M_\sun$ (magenta), $1.35~M_\sun$ (blue),
and $1.37~M_\sun$ (black) WDs.
(c) Neon nova 2 (Ne2): $1.3~M_\sun$ (red), $1.33~M_\sun$ (magenta),
$1.35~M_\sun$ (blue), and $1.37~M_\sun$ (black) WDs.
(d) Neon nova 3 (Ne3): $1.3~M_\sun$ (red), $1.33~M_\sun$ (magenta),
$1.35~M_\sun$ (blue), and $1.37~M_\sun$ (black) WDs.   
\label{individual_v1674_her_v_chemical_logscale_no2}}
\end{figure*}

\subsection{Model A and model B: comparison of different mass-accretion rate
models} 

We present two theoretical nova models for a 1.35 $M_\sun$ WD
with different mass accretion rates;
 Model A ($1 \times 10^{-11}~M_\odot$ yr$^{-1}$) and model B
($5 \times 10^{-10}~M_\odot$ yr$^{-1}$).

As shown in Table \ref{table_models},
model A has a larger ignition mass, that yields
a higher maximum temperature $T^{\rm max}$ and a
larger maximum luminosity at ignition $L_{\rm nuc}^{\rm max}$.
Thus the shell flash is stronger in model A.
As a result, it shows a faster evolution toward the maximum expansion,
i.e., the shorter duration of X-ray flash (Figure \ref{xrayflash}) and
faster rising speed in optical (Figure \ref{light.2}).

With a larger envelope mass, model A reaches a larger maximum expansion
of the photosphere at stage G as in Figure \ref{hr}.
The wind mass loss rate, which attains
the maximum at the maximum expansion of the photosphere,
is also larger than model B.
This results in the difference of the $V$ brightness. Model A is brighter by
0.64 magnitudes than model B, because the luminosity depends strongly
on the wind mass loss rate (equation (\ref{free-free_flux_v-band})).

\begin{figure}
\epsscale{1.15}
\plotone{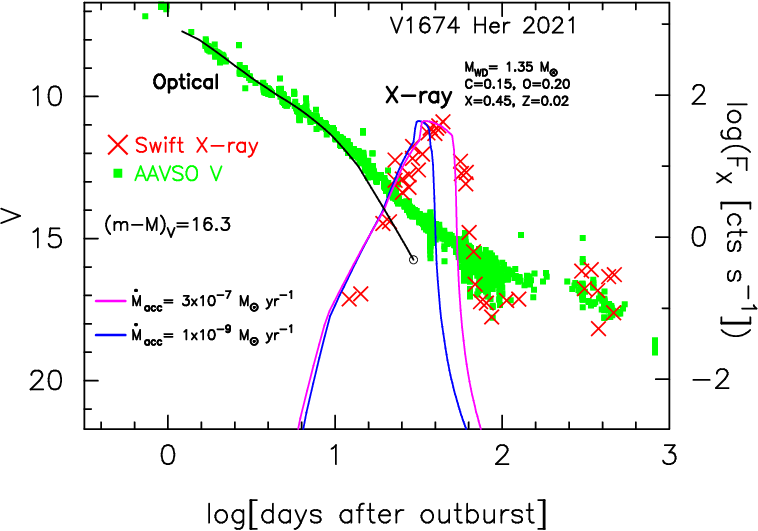}
\caption{
Our model $V$ and X-ray light curves for V1674 Her as well as the observed
$V$ and X-ray light curves.  The model has the $1.35 ~M_\sun$ WD mass
for the chemical composition of CO3 (same as that in Figure
\ref{individual_v1674_her_v_chemical_logscale_no2}b).
Two mass-accretion rates are assumed during the nova outburst, i.e.,
$\dot{M}_{\rm acc}= 1\times 10^{-9}$ (blue line for X-ray)
and $3\times 10^{-7}$ (magenta line) $ ~M_\sun$ yr$^{-1}$.
A high mass-accretion rate makes the X-ray turnoff date later
on day 60 or more.
\label{v1674_her_v_x45z02c15o20_large_dotM_logscale}}
\end{figure}

On the other hand, in the post-maximum phase,
model A and model B show very similar light curves.
In post-maximum stages a nova envelope almost settles down
to a steady state structure because nuclear burning energy generation
rate has decreased to be comparable with the photospheric
luminosity ($L_{\rm nuc}\approx L_{\rm ph}$ in Figure \ref{Levo}a).
Thus, the light curves of model A and
model B converge in Figure \ref{light.2}.
After the wind stops, the novae enter the SSS phase, the hydrogen-rich
envelopes of which are also very similar between model A and model B
(see Figure \ref{TRV.compari}).

To summarize, these theoretical nova evolution/light curves depend
strongly on the mass accretion rate in the rising (pre-maximum) phase,
but are independent of the mass accretion rate in the decay (post-maximum) phase.

\citet{kat83} and \citet{kat94h} presented a hypothesis that nova decay phase
can be well followed with a steady state sequence independently
of the mass-accretion rate.
The good agreement in the light curves among model A, model B, and
the steady-state model (blue dotted line) in Figure \ref{TRV.compari}
confirms that this hypothesis is reasonable.

\subsection{Comparison with steady-state sequence}\label{sec_steadystate}



In what follows, we discuss more about the light curve of V1674 Her
based on the nova light curve models of steady-state sequences  
\citep{kat94h}.

\subsubsection{Applicability of steady-state model light curves}
\label{sslightcurve}

We have calculated a sequence of steady state wind solutions 
for a 1.35 $M_\sun$ WD with the chemical composition of $X=0.55$, 
$Y=0.3$, $Z=0.02$, and 
$X_{\rm CNO}=0.13$, where $X_{\rm CNO}$ is the additional 
carbon, nitrogen, and oxygen other than $Z=0.02$. 
$Z=0.02$ contains all the heavy elements with solar composition
including C, N, and O.
The radius at the bottom of the envelope (i.e., nuclear burning region)
to be $\log R/R_\sun=-2.48$. 
These parameters are taken from model A in the decay phase. 
Model B shows very similar composition.  

Figure \ref{TRV.compari}a depicts the free-free emission light curve 
of this steady state sequence (dotted blue line). 
We see a good agreement between the steady state sequence 
and time-dependent model over 20 days from the optical peak.  
This agreement confirms that 
the nova evolution can be followed 
by a sequence of steady-state solutions \citep{kat83,kat94h}. 
Also the good agreement in the light curves of model A and model B 
demonstrates that the decay phase of a nova light curve is independent 
of the mass-accretion rate.

\subsubsection{Parameter dependence of steady-state light curve}

In the light curve fitting, we chose a 1.35 $M_\odot$ WD for V1674 Her. 
The core-material mixing is assumed to be $\Delta X_{\rm C}=0.1$ at 
ignition. 
The resultant envelope composition is $X=0.55$, $Y=0.29$, 
$Z=0.02$, $X_{\rm CNO}=0.14$. 
for both model A and model B. 
If we adopt different WD mass or different way of 
core mixing, we may have different light curves.   
Here we examine how the light curve depends on the WD mass 
and composition, using our database of steady-state light curves. 

We have published many steady-state light curves 
for different WD masses and chemical compositions 
\citep[e.g.,][]{kat94h, hac25kv392per}. 
Among our database, we plot four WD masses 
of 1.37, 1.35, 1.33, and $1.3 ~M_\sun$ with four typical chemical 
compositions (a:CO2, b:CO3, c:Ne2, and d:Ne3) in Figure
\ref{individual_v1674_her_v_chemical_logscale_no2}.
The exact composition is denoted in each panel. 
For the same chemical composition, a more massive WD shows more 
rapid decline in the $V$ magnitude and earlier X-ray turn-on time. 
We exclude the 1.3 and 1.33 $M_\sun$ models that show 
slower $V$ light curves than observational data in 
all of the four compositions.   
Also the 1.37 $M_\sun$ WDs show earlier decay light curves and 
can be excluded. 

These figures also show the X-ray count rates obtained with 
the Swift/XRT. Note that these are count rates, which are 
different from the X-ray absorbed flux in Figure \ref{TRV.compari}a. 
The fitting in the vertical direction of the X-ray light curve 
is arbitrarily: vertically shifted to fit the maximum X-ray count rate.     

Among these models, the $1.35 ~M_\sun$ WD (CO3) in Figure
\ref{individual_v1674_her_v_chemical_logscale_no2}b best follows
both the $V$ light curve as well as the rising phase of the 
X-ray turn-on time. This model has CO3 composition, which 
is close to the chemical composition of model A and model B.  
Thus, our choice of $1.35 ~M_\sun$ WD and the way of 
core material mixing are reasonable. 

However, this model shows that the X-ray turnoff time is too early, 
i.e., the SSS phase is too short. 
All the other theoretical models also show too short SSS phases.  
This problem will be discussed in the next subsection.

\subsection{Long SSS phases driven by high mass-accretion rates}
\label{acc_rate_postnova}

The SSS phase of V1674 Her 
starts from day $\sim 20$ and ends on day $\sim 60$. 
The duration is $\sim 40$ days which is much
longer than the $\sim 10$ day duration in our model A and model B, 
and all the 1.35 $M_\sun$ models in 
Figure \ref{individual_v1674_her_v_chemical_logscale_no2}. 

This discrepancy could be solved if the mass accretion 
from the companion star restarts/continues after the wind stops. 
During the SSS phase, the companion star has been strongly 
irradiated and heated by the hydrogen burning WD, 
thus, the mass-accretion rate could be enhanced. 
If the mass accretion rate is as high as the hydrogen burning rate
(mass consumption rate of hydrogen-rich envelope) 
but slightly lower than that, the SSS phase 
lasts longer because new fuel is supplied.
Such a trend of enhanced mass-accretion rate has already been confirmed
in the $V$ light curve analysis in 
KT Eri \citep{hac25kw} and in V392 Per \citep{hac25kv392per}. 
The orbital periods of these novae are rather long
(2.6 days and 3.2 days, respectively) and, therefore,
they have a very large disk and the viscous heated disk is bright.

To show this effect, we assume a high rate of 
$\dot{M}_{\rm acc}=3\times 10^{-7} ~M_\sun$ yr$^{-1}$ during the outburst.
The resultant X-ray light curve 
is shown in Figure \ref{v1674_her_v_x45z02c15o20_large_dotM_logscale} for
a $1.35 ~M_\sun$ WD (CO3). Comparing with the mass accretion rate 
of $\dot{M}_{\rm acc}=1\times 10^{-9} ~M_\sun$ yr$^{-1}$,
the X-ray turn-on time is hardly changed,
but the end day of hydrogen burning is prolonged
up to day 60, being roughly consistent with the Swift/XRT observation.



\begin{figure}
\gridline{\fig{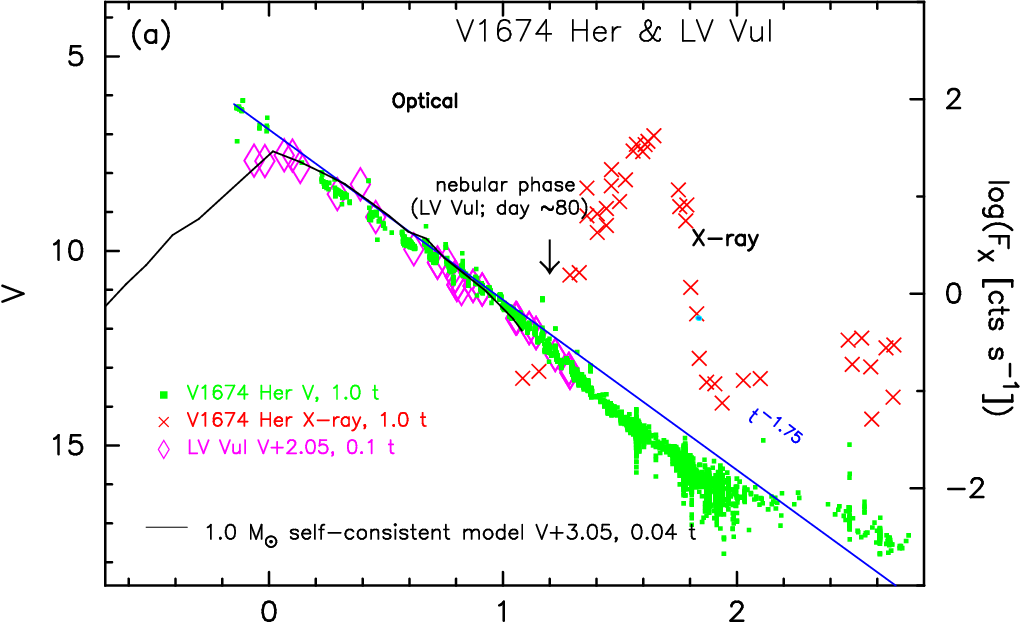}{0.45\textwidth}{}
          }
\gridline{
          \fig{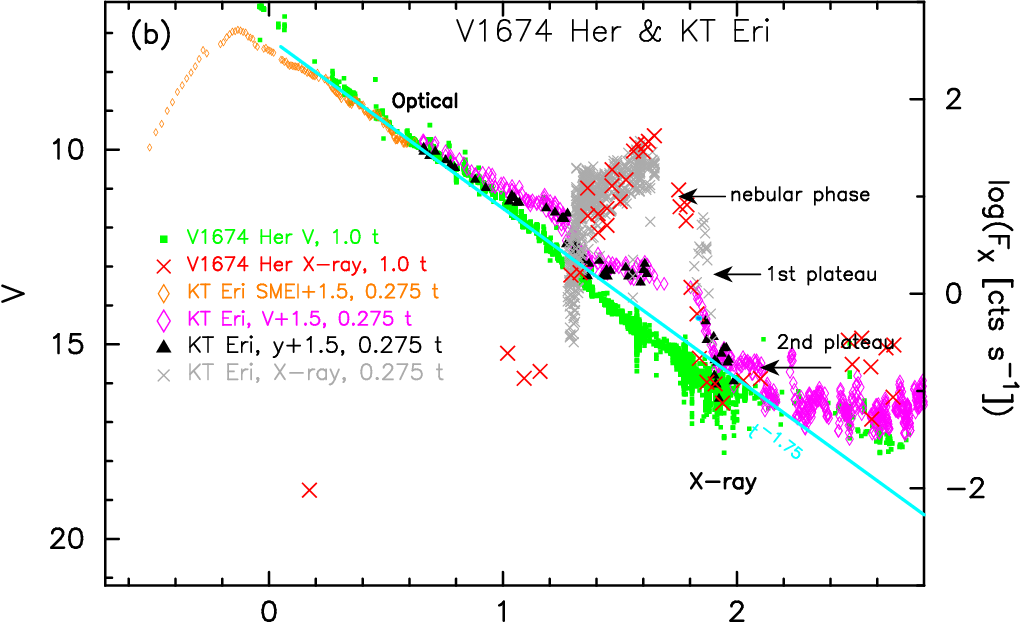}{0.45\textwidth}{}
          }
\gridline{
          \fig{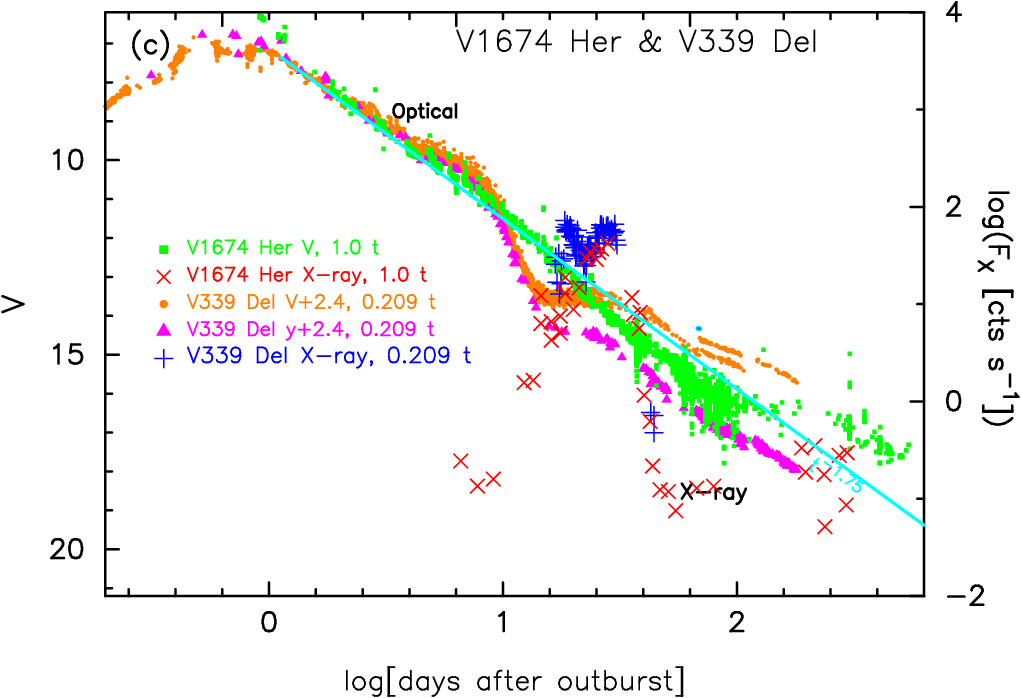}{0.45\textwidth}{}
          }
\caption{
(a) Two $V$ light curves of V1674 Her and LV Vul are
overlapped along Equation (\ref{overlap_brigheness}).
The text ``LV Vul V+2.05, 0.1 t'', for example, means $f_{\rm s}=0.1$ and
$\Delta V=+2.05$, for the template nova LV Vul against the $V$ light curve
of the target nova V1674 Her (``V1674 Her V, 1.0 t'').
We also add \citet{kat22sha}'s fully self-consistent nova model (thin black
line). 
(b) Same as panel (a), but for the template nova KT Eri.
(c) Same as panel (a), but for the template nova V339 Del.
The data of LV Vul, KT Eri, and V339 Del are the same as those
in \citet{hac23k}, \citet{hac25kw}, and \citet{hac24km}, respectively.
\label{v1674_her_kt_eri_v339_del_v_x_stretching}}
\end{figure}


\section{Conclusions}\label{sec_conclusion}

Our results are summarized as follows: \\
\begin{enumerate}
\item Our theoretical free-free emission light curve models
for a 1.35 $M_\sun$ WD reasonably reproduce the observed optical
light curve from the very early rising phase to the later phase. 
Model A ($\dot{M}_{\rm acc}=1 \times 10^{-11}~M_\odot$ yr$^{-1}$)
shows a 0.94 magnitudes brighter peak in the $V$ band than in model B
($\dot{M}_{\rm acc}=5 \times 10^{-10}~M_\odot$ yr$^{-1}$). 
\item 
The temporal variation of our model photospheric radius
is consistent with the observed epochs on both appearances of
the orbital and spin flux modulations, respectively.
\item We obtained the $V$ band distance modulus of 
$(m-M)_V= 16.3\pm 0.2$ from the time-stretching method of nova light curves. 
Also the distance to V1674 Her is determined to be $d=8.9\pm 1$ kpc 
together with the interstellar extinction of $E(B-V)= 0.5 \pm 0.05$ in 
Figure \ref{distance_reddening_v1674_her}.
\item Model A and model B show very similar light curves 
in the post-maximum phase. Also these light curves agree well
with the light curve of the sequence of 
steady state wind solutions. 
This confirms that the decay phase of a nova outburst can be well 
approximated by the sequence of steady wind mass-loss solutions,  
independently of the mass accretion rate. 
\item Among our database of steady-state light curves 
for various WD masses and chemical compositions of envelope,   
the best-fit model is a $1.35 M_\sun$ WD with the chemical
composition of CO3.  
This confirms that our choice of a $1.35 ~M_\sun$ WD is reasonable.
\item The observed SSS phase is much longer than those predicted by 
models A and B.  This discrepancy can be solved/relaxed
if we assume high mass accretion rates such as 
$\dot{M}_{\rm acc}=3\times 10^{-7} ~M_\sun$ yr$^{-1}$ during the SSS phase 
(Figure \ref{v1674_her_v_x45z02c15o20_large_dotM_logscale}).

\item Our shock formation model based on model A
predicts that a strong shock arises just after the maximum
expansion of the photosphere (on day 0.32).
This epoch is consistent with the first detection of GeV gamma-rays
with the Fermi/LAT on day $0.39\pm 0.25$.

\item This model also gives a reasonable explanation on the 
observed properties of the shocked shell in the ejecta, such as 
the temperature, column density, and time variability of hard X-rays,
and the temporal change of radio luminosity.

\end{enumerate}

\begin{acknowledgments}
We thank the American Association of Variable Star Observers
(AAVSO) for the archival data of V1674 Her.
We are also grateful to the anonymous referee for useful comments 
that improved the manuscript. 
\end{acknowledgments}

\vspace{5mm}
\facilities{Swift(XRT), AAVSO}


\appendix

\section{Time-stretching Method}
\label{time_stretching_method}


The time-stretching method is a powerful way
to obtain the distance modulus in the $V$ band, $(m-M)_V$,
toward a nova, which has ever been applied to many novae  
\citep{hac10k, hac15k, hac16k, hac18k,hac20skhs}. 
In this appendix, we explain the time-stretching method and
determine the distance modulus to V1674 Her.

\subsection{Time-Stretching Method}
\label{sec_timestretching}

Nova light curves often show a common property; if two nova light curves 
are plotted in the logarithmic time and shift vertical and horizontal 
directions, the major part of these light curves are overlapped 
each other independent of the WD mass, chemical composition, and speed 
class of novae \citep{hac06kb, hac20skhs}. 
Using this remarkable property, we can determine 
the distance to a nova (target nova: V1674 Her) by comparing 
a well studied nova with known distance (template nova). 

Here, we describe the $V$ light curves of the target nova as 
$(m[t])_{V,\rm target}$ and the template nova $(m[t])_{V,\rm template}$. 
When we adopt an appropriate time-stretching parameter $f_{\rm s}$, 
these two nova $V$ light curves overlap each other.  
We shift the target nova light curve in the horizontal direction 
by a factor of $f_{\rm s}$ in the logarithmic scale 
($t \rightarrow t\times f_{\rm s}$), 
and move vertically down by $\Delta V$. This vertical shift 
can be written as 
\begin{equation}
(m[t])_{V,\rm target} = \left((m[t \times f_{\rm s}])_V
+ \Delta V\right)_{\rm template}. 
\label{overlap_brigheness}
\end{equation}
As the two nova light curves are overlapping, 
their distance moduli in the $V$ band satisfy
\begin{eqnarray}
& & (m-M)_{V,\rm target} \cr
&=& ( (m-M)_V + \Delta V )_{\rm template} - 2.5 \log f_{\rm s}.
\label{distance_modulus_formula}
\end{eqnarray}
Here, $m_V$ and $M_V$ are the apparent and absolute $V$ magnitudes,
and $(m-M)_{V, \rm target}$ and $(m-M)_{V, \rm template}$ are
the distance moduli in the $V$ band
of the target and template novae, respectively.
\citet{hac18k, hac18kb, hac19k, hac19kb, hac21k} confirmed that
Equations (\ref{overlap_brigheness}) and (\ref{distance_modulus_formula})
are also broadly valid for other $U$, $B$, and $I$ (or $I_{\rm C}$) bands.

\subsection{Distance Modulus in the $V$ Band}
\label{distance_v_band} 

This remarkable similarity is demonstrated in
Figure \ref{v1674_her_kt_eri_v339_del_v_x_stretching}, which
compares the $V$ and X-ray light curves of V1674 Her with (a) LV Vul,
(b) KT Eri, and (c) V339 Del.
These novae decline much slower, 
but are time-stretched into almost one line in the figure.
Here, the outburst day of V1674 Her is assumed to be $t_{\rm OB}=$
JD 2,459,377.68 (UT 2021 July 12.18).

In Figure \ref{v1674_her_kt_eri_v339_del_v_x_stretching}a,
we regard V1674 Her as the target and LV Vul as the template
in Equation (\ref{overlap_brigheness}). 
As LV Vul evolves 10 times slower, we adopt $f_{\rm s}= 0.1$
and $\Delta V= +2.05$ and have the relation of
\begin{eqnarray}
(m&-&M)_{V, \rm V1674~Her} \cr
& = & (m - M + \Delta V)_{V, \rm LV~Vul} - 2.5 \log 0.1 \cr
&=& 11.85 + 2.05\pm 0.2 + 2.5 = 16.4\pm 0.2,
\label{distance_modulus_v1674_her_lv_vul_v}
\end{eqnarray}
where we adopt $(m-M)_{V, \rm LV~Vul}=11.85$ from \citet{hac18k}.


Similarly we apply our time-stretching method to a pair of 
V1674 Her and KT Eri, and a pair of V1674 Her and V339 Del,
and plot them in Figure \ref{v1674_her_kt_eri_v339_del_v_x_stretching}b
and c, respectively.
Then, we have
\begin{eqnarray}
(m&-&M)_{V, \rm V1674~Her} \cr
&=& ((m - M)_V + \Delta V)_{\rm KT~Eri} - 2.5 \log 0.275 \cr
&=& 13.4 + 1.5\pm0.2 + 1.4 = 16.3\pm0.2 \cr
&=& ((m - M)_V + \Delta V)_{\rm V339~Del} - 2.5 \log 0.209 \cr
&=& 12.2 + 2.4\pm0.2 + 1.7 = 16.3\pm0.2,
\label{distance_modulus_v_v1535_sco_v339_del_kt_eri}
\end{eqnarray}
where we adopt $(m-M)_{V, \rm KT~Eri}=13.4$ from \citet{hac25kw} and
$(m-M)_{V, \rm V339~Del}= 12.2$ from \citet{hac24km}.
Thus, we obtain $(m-M)_{V, \rm V1674~Her}= 16.33\pm 0.2$.

\end{document}